\documentclass[a4paper,11pt]{article}
\pdfoutput=1 

\usepackage{jcappub} 

\usepackage[T1]{fontenc} 
\usepackage[utf8]{inputenc}
\usepackage{bm}
\usepackage{aas_macros}
\usepackage{multirow}
\usepackage{xcolor}

\title{\boldmath Cosmology of Single Species Hidden Dark Matter}

\author[a,b,c,1]{Weikang Lin,\note{Corresponding author.}}
\author[d]{Xingang Chen,}
\author[e]{Himanish Ganjoo,}
\author[c,f,g,h]{Liqiang Hou,}
\author[g,h,i]{Katherine J. Mack}

\affiliation[a]{South-Western Institute For Astronomy Research, Yunnan University, Kunming 650500, Yunnan, P. R. China}
\affiliation[b]{Tsung-Dao Lee Institute, Shanghai Jiao Tong University, Shanghai 200240, China}
\affiliation[c]{Department of Physics, North Carolina State University, Raleigh, NC 27695, USA}
\affiliation[d]{Institute for Theory and Computation, Harvard-Smithsonian Center for Astrophysics, Cambridge, MA 02138, USA}
\affiliation[e]{Laboratoire d’étude de l’Univers et des phénomènes eXtrêmes, Université Paris Cité, Observatoire de Paris, Université PSL, CNRS, F-92190 Meudon, France}
\affiliation[f]{Institute for Advanced Study in Physics, Zhejiang University, Hangzhou, 310058 China}
\affiliation[g]{Perimeter Institute for Theoretical Physics,
Waterloo, Ontario, N2L 2Y5, Canada}
\affiliation[h]{Department of Physics and Astronomy, University of Waterloo, Waterloo, ON N2L 3G1, Canada}
\affiliation[i]{Department of Physics and Astronomy, York University, Toronto, Ontario, M3J 1P3, Canada}

\emailAdd{weikanglin@ynu.edu.cn, xingang.chen@cfa.harvard.edu, Himanish.Ganjoo@obspm.fr, lhou2@ncsu.edu, kmack@perimeterinstitute.ca}

\abstract{Cosmology and astrophysics provide various ways to study the properties of dark matter even if they have negligible non-gravitational interactions with the Standard Model particles and remain hidden. We study a type of hidden dark matter model in which the dark matter is completely decoupled from the Standard Model sector except gravitationally, and consists of a single species with conserved comoving particle number and conserved comoving entropy. This category of hidden dark matter includes models that act as warm dark matter but is more general. In particular, in addition to having an independent temperature from the Standard Model sector, it includes cases in which dark matter is in its own kinetic equilibrium or is free-streaming, obeys fermionic or bosonic statistics, and processes a chemical potential that controls the particle occupation number. While the usual parameterization using the free-streaming scale or the particle mass no longer applies, we show that all cases can be well approximated by a set of functions parameterized by only one parameter as long as the chemical potential is nonpositive: the characteristic scale factor at the time of the relativistic-to-nonrelativistic transition. We study the constraints from Big Bang Nucleosynthesis, the cosmic microwave background, the Lyman-$\alpha$ forest, and the smallest halo mass. We show that the most significant phenomenological impact is the suppression of the small-scale matter power spectrum -- a typical feature when the dark matter has a velocity dispersion or pressure at early times. So far, the Lyman-$\alpha$ forest and the small dark matter halo population provide the strongest constraints, limiting the transition redshift to be larger than $\sim6.2\times10^7$.}

\begin{document}
\maketitle
\flushbottom

\section{Introduction}
The dark matter hypothesis consistently explains a wide range of cosmological and astrophysical observations. The first of these lines of evidence came from the unexpected observation of high galactic velocity dispersion in clusters \cite{zwicky1933spectral,smith1936Virgomass}. Later, rotation curves in spiral galaxies were found to be flat in the outer region \cite{1957-van_de_Hulst}, which cannot be explained by only visible matter with Newtonian gravity. Observations of the abundance of the primordial elements created during Big Bang Nucleosynthesis (BBN) suggest that dark matter is non-baryonic \cite{2003BBN}. Cosmological observations across the cosmic microwave background (CMB) \cite{Planck2018-parameter-constraints} and large scale structure (LSS) \cite{DES-y1-2018-joint,Hikage-etal-2019,Joudaki-etal-2018} are well fitted by the standard cosmological model, which assumes cold dark matter. In addition, the recent discovery of the Bullet Cluster (in which the lensing and the X-ray images separate from each other \cite{bulletcluster-NASA}) and the significant drop in the correlation between stellar velocity dispersion and luminosity for ultra-faint dwarf galaxies \cite{Lin:2016vmm} both favor the dark matter hypothesis over the modified gravity/Newtonian dynamics hypothesis. 

Efforts to identify the nature of the dark matter particles, on the other hand, turn out to be difficult. Studies have been conducted to investigate the impacts on and constraints from cosmological and astrophysical observations of proposed dark matter properties. These include, but are not limited to, dark matter annihilation \cite{HESS-DM-anni-2018,AMS-DM-anni-2014,Chan-etal-2019-MD-anni,Depta-etal-2019,Boudaud-etal-2018-DM-constraint,Schon-etal-2018,Schon-etal-2015}, decay \cite{Poulin-etal-2016,Enqvist-etal-2015-DM-decay,Blackadder-Koushiappas-2016}, interaction with baryons \cite{Dvorkin-Blum-Kamionkowski-2014-DM-B-interaction,Visinelli:2015eka,Xu-Dvorkin-Chael-2018-DM-B-interaction,Kimberly-etal-2018-DM-B-interaction,Ooba-etal-2019-DM-B-interaction,2018-Barkana-eta-not-DM-B-interaction,Nadler-2019-subhalo-DM-B-v-independent-interaction} (also see constraints from direct detection \cite{Adhikari-etal-2019,Beck-etal-2019,Cresst-2017-DM-B-interaction,Lux-2017-DM-B-interaction,Xenon-2016-DM-B-interaction,Xenon-2017-DM-B-interaction}), and temperature \cite{Carlson-etal1992,Bode-2001,Viel-etal-PhysRevLett2008,Safarzadeh-etal-2018-WDM-21cm,Boyarsky-etal-2019-WDM-21cm}. So far, only increasingly strong upper bounds of those properties have been obtained \cite{BhupalDev:2013oiy}, and a wide region of the parameter space of dark matter models has been ruled out. This, however, does not falsify the existence of dark matter, but instead motivates the quest for new models \cite{Garcia-Kaneta-Mambrini-2020,2020-Erickcek-Ralegankar-Shelton,2019-Miller-Erickcek-Murgia,Blanco:2019eij,Bertone-Tait-2018-DM-summary,Battaglieri-2017-DM-newideas}.

A common starting assumption in dark matter models is that the dark matter was at some point in thermal equilibrium with Standard Model particles, or at least had some appreciable thermal contact \cite{Lee-Weinberg-PhysRevLett-1977,Dicus-etal-PhysRevLett-1977,Dodelson-Widrow-1994,Colombi-Scott-Widrow-1996,Viel-etal-2005,Bode-2001}; this is required for the so-called ``WIMP Miracle''. However, in this paper, we will abandon this assumption and consider the case in which the dark matter has a totally independent thermal history from the Standard Model sector. Such a scenario may be naturally realized in, for example, string theory-motivated inflation and reheating models in warped compactification. In this scenario, the inflaton decays to particles residing in multiple sectors, and the dark matter resides in a sector whose non-gravitational couplings to the Standard Model sector are negligible and is thus ``hidden'' \cite{Chen-Tye-2006}.\footnote{The dark matter can also effectively decouple from the standard model in the same sector if, as a decay product of the inflaton, its mass exceeds the reheat temperature \cite{Allahverdi:2002nb,Allahverdi:2002pu}.} For the purpose of this paper, we will only be interested in the phenomenological properties of the hidden dark matter after it is generated, so the reheating scenarios and the UV-completed construction of the hidden sector will henceforth not be important. By removing the direct coupling between the dark matter and visible sector, hidden dark matter models provide unconventional viewpoints on the dark matter phenomenology, and may be used to explain some existing anomalies  \cite{Chen-Tye-2006,Foot:2014osa,Foot:2014uba,Foot:2016wvj,Blanco:2019eij2015-Erickcek,Li:2023nez}. We will show that even if the direct interactions between dark matter and Standard Model particles are too weak to generate direct detection signals or other cosmological effects, cosmological and astrophysical phenomena will still allow us to probe some cosmological evolutionary properties of such dark matter due to the gravitational coupling.

A wide range of DM models can be found in the literature: e.g, a constant DM equation of state \cite{2005-Muller-constraint-on-wDM,2009-Calabrese-etal-CMB-on-WDM}, binned equation of state \cite{Kopp-etal-2018PhRvL}, ballistic dark matter \cite{2018Das-etal-Ballistic-DM}, or a reduced relativistic dark matter model in which all dark matter particles have the same velocity \cite{2018-Hipolito-Ricaldi-etal,2005-DeBerredo-etal-reduced-relativistic-DM}. Additionally, 
extensive studies of generalized dark matter (GDM) models can also be found in Refs.\,\cite{Hu1998ApJ,Kopp-etal-2016-GDM,Kopp-etal-2018PhRvL,Kumar-etal-2019-WDM}. Those models are not necessarily based on the particle nature of dark matter, which makes the connection between the equation of state and the particle properties of dark matter less apparent, and it is difficult to extract information (e.g., the initial velocity dispersion) for cosmological N-body simulations. We aim to study a hidden dark matter (HiDM) model\footnote{Note that ``Hi'' here stands for ``hidden'', to distinguish from hot dark matter (HDM). The highly decoupled dark matter may also have relatively weak interactions with the Standard Model sector, which generates interesting phenomenological consequences \cite{Feldman:2006wd,Feldman:2007nf,Feng:2008ya,Feng:2008mu,Chen:2009iua,Lin:2022xbu,Lin:2022mqe,Lin:2022niw}. In this paper we will study the limit in which non-gravitational interactions with the Standard Model are ignored, and the term ``hidden" is used to specifically denote this limit.} that is theoretically consistent with some underlying particle properties, and constrain it from a wide range of cosmological and astrophysical observations. Another motivation is to determine whether such a nonstandard dark matter scenario can resolve some current problems in cosmology, e.g., the $H_0$ and the $\sigma_8$ tensions \cite{2019-Riess-etal,Joudaki-etal-2018}; but also see \cite{Lin:2019htv,Lin:2021sfs}.

While allowing an independent thermal history, among several possibilities discussed in Ref.~\cite{Chen-Tye-2006}, we will only consider the simplest situation in which the one-species HiDM has the following properties:
\begin{itemize}
    \item The dark matter's \emph{entropy per unit comoving volume is a constant.}
    \item The dark matter's \emph{particle number per unit comoving volume is a constant.}
\end{itemize}
Together, these two properties imply that the \emph{specific entropy is a constant}. Assuming HiDM starts with its own kinetic equilibrium phase, we will consider two cases. \textbf{Case 1}: The HiDM is in kinetic equilibrium at all times with a temperature different from that of the Standard Model sector. \textbf{Case 2}: The HiDM started in kinetic equilibrium with a temperature different from that of the Standard Model sector but later became free-streaming during its relativistic phase. These are two extreme cases corresponding to the strong and weak self-interaction limits, respectively. An intermediate case is where the HiDM maintains its own kinetic equilibrium until it becomes free-streaming at some time during the relativistic-to-nonrelativistic transition. 
We show in this work that the two extreme cases are phenomenologically very similar and their background evolution can be well represented by one universal parameterized function of the scale factor $a$. Therefore, we expect that the intermediate case should also behave similarly.  

Note that we do not have any assumption on the HiDM particle's mass, number of intrinsic degrees of freedom, and whether or not it was in thermal contact with Standard Model particles in the past. HiDM has some similar properties but is more general compared to warm dark matter (WDM) \cite{Bond-Efstathiou-Silk-1980-PhysRevLett,Viel-etal-2005,Viel-etal-PhysRevLett2008,2013-Schneider-etal,Colombi-Scott-Widrow-1996}. We summarize the differences as follows
\begin{itemize}
    \item We consider both fermionic and bosonic statistics.
    \item We include a chemical potential to control particle occupation number.
    \item We consider that the HiDM can be in its own kinetic equilibrium or free streaming.
\end{itemize}
We dub the hidden dark matter cosmology $\Lambda$HiDM. These differences, especially the inclusion of chemical potential, modify several physical relations established in the WDM case and motivate us to look for another parameter different from the usually adopted free-streaming scale or the particle mass.

We organize our paper as follows: In Sec.\,\ref{section:background-level} we describe the background evolution of the HiDM model and estimate a constraint from BBN. We show that the HiDM model can be well parameterized by just the relativistic-to-nonrelativistic transition time, and cosmological phenomenology is insensitive to other particle properties. In Sec.\,\ref{section:perturbation}, we describe the scalar perturbations in the HiDM model. In Secs.\,\ref{section:observational-impacts-and-observations-CMB}, \ref{section:observational-impacts-matter-power} and \ref{section:HiDM-nonlinear-scales}, we individually study phenomenology at different scales corresponding to different observations and derive their constraints separately. We then list the constraints in Sec.\,\ref{section:list-of-constraints} and summarize in Sec.\,\ref{section:HiDM-summary}. Throughout this work, we use natural units with $c=k_{\rm{B}}=1$. When needed, we will take the standard $\Lambda$CDM model with parameters given by mean values of the Planck 2018 TTTEEE+lensing+BAO constraints \cite{Planck2018-parameter-constraints}.

\section{Background evolution and similarities between different cases}\label{section:background-level}
In this section, we present the background evolution of the HiDM scenario. Properties of different cases are summarized in Table \ref{tab:different-cases}. At the end of this section, we estimate an observational constraint on HiDM from BBN. 

\begin{table}[]
\footnotesize
    \centering
    \begin{tabular}{lcccc}
    \hline
    \hline
           Case   & Distribution & Statistics & Chemical potential & Perturbation \\
           \hline
           (I) Kinetic equilibrium   & Eq.~\eqref{eq:HiDM-distribution} & \multirow{2}{*}{Fermion/boson} & \multirow{2}{*}{$\mu_{\rm ini}\leq0$} & Without pressure anisotropy \\
           (II) Free-streaming  & Eq.~\eqref{eq:HiDM-free-stream-p-distribution}  & &  & With pressure anisotropy \\
           \hline
    \end{tabular}
    \caption{Different cases considered in the single-species HiDM model. Both fermion and boson cases are considered. The classical case is defined as $\mu_{\rm ini}\rightarrow-\infty$.}
    \label{tab:different-cases}
\end{table}

\subsection{Numerical solutions to the background evolution}\label{section:numerical-solution-BG-short}
\subsubsection{Background evolution I - kinetic equilibrium case}\label{sec:background-equilibrium}
We first consider the HiDM in the kinetic equilibrium case. Given a mass $m_{\rm h}$, number of intrinsic degrees of freedom $g$, temperature $T_{\rm h}$, and chemical potential $\mu_{\rm h}$, the physical momentum distribution in kinetic equilibrium is given by
\begin{equation}\label{eq:HiDM-distribution}
    f_{\rm{eq}}(p;T_{\rm h},\mu_{\rm h}) = \frac{g}{2\pi^2\hbar^3}\frac{p^2}{\exp\big[(\sqrt{p^2+m_{\rm h}^2}-\mu_{\rm h})/T_{\rm h}\big]\pm1}\,,
\end{equation}
with $+1$ in the denominator for fermions and $-1$ for bosons. The subscript ``h'' refers to the hidden sector. Since $T_{\rm h}\propto 1/a$ in the relativistic phase where $a$ is the scale factor normalized to be $1$ today, we define $T_{\rm h}^0\equiv\lim_{a\rightarrow0}aT_{\rm h}$ and the initial ``warmness'' as
\begin{equation}\label{eq:initial-warmness}
    {\text{Initial Warmness}\equiv T_{\rm h}^0/m}\,.
\end{equation}
The chemical potential here is not associated with the number of dark matter particles minus that of the anti-dark matter particles but with the total dark matter number. The chemical potential is usually considered to be zero, especially in the case where dark matter annihilates with anti-dark matter into photons. Since we consider a HiDM that is decoupled from the Standard Model sector, we allow a non-zero chemical potential. To simplify the calculation, we define some dimensionless quantities: $x\equiv m_{\rm h}/T_{\rm h}$, $y\equiv \mu/T_{\rm h}$, $z\equiv p/T_{\rm h}$, and $\Delta\equiv y-x=\frac{\mu_{\rm h}-m_{\rm h}}{T_{\rm h}}$. The physical meaning of $\Delta$ is the effective chemical potential-to-temperature ratio. We call $\mu_{\rm h}-m_{\rm h}$ the effective chemical potential because 1. $m_{\rm h}$ can be ignored at the relativistic limit and 2. the nonrelativistic limit of Eq.~\eqref{eq:HiDM-distribution} reduces to,
\begin{equation}\label{eq:HiDM-distribution-non-relativistic}
    f(p; x\equiv\frac{m_{\rm{h}}}{T_{\rm{h}}}\rightarrow\infty) = \frac{g}{2\pi^2\hbar^3}\frac{p^2}{\exp\big[(\frac{p^2}{2m_{\rm{h}}}-\mu_{\rm{h}}+m_{\rm{h}})/T_{\rm{h}}\big]\pm1}\,.
\end{equation}
We can see that it is $\mu_{\rm{h}}-m_{\rm{h}}$ that serves as the effective chemical potential. When $\Delta\ll-1$, the distribution reduces to the classical case. Quantum statistic effects take place when $\Delta\gtrsim-1$. For fermions, these effects manifest as a degenerate pressure, which is larger for larger $\Delta$. For bosons, the pressure will be smaller for larger $\Delta$. In this work, we restrict ourselves to the case that $\Delta\leq0$ for both fermion and boson cases. A strong degenerate pressure will be present for fermions when $\Delta > 0$, and Bose-Einstein condensation will occur for bosons when $\Delta = 0$; the study of these topics is left for another work \cite{Zhang:2024vkx}.

We shall show that $\Delta$ is a constant when HiDM is relativistic, so we denote $\Delta_{\rm ini}$ as the initial value of $\Delta$. A nonzero $\Delta$ has a similar (but not the same) role to that of the normalization factor $\beta$ in the warm dark matter model considered in \cite{Colombi-Scott-Widrow-1996}, because, in the classical limit, $\exp(\Delta)$ becomes the normalization factor of the distribution. But, our $\Delta$ is more physically motivated and specifies the particle occupation number. Species considered in standard cosmology usually have $\Delta_{\rm{ini}}=0$. However, we release this restriction and only require it to be non-positive. We define the classical case as the one with $\Delta_{\rm{ini}}\ll-1$. The fiducial case has $\Delta_{\rm ini}=0$. The cases with $\Delta_{\rm ini}<0$ correspond to those where DM is produced with a smaller number density than the standard case for a given initial warmness. This may occur, e.g., when HiDM was generated at the end of brane inflation with an arbitrary temperature and number density \cite{Chen-Tye-2006}. Or, it was produced by decays of the inflaton, other particles, or cosmic string with a small rate (like the freeze-in mechanism \cite{BhupalDev:2013oiy}) and then attain kinetic equilibrium via some self-interaction. We note that cannibalistic interactions can change the comoving number density even for one species DM \cite{Pappadopulo:2016pkp}, which we do not consider.

The process of solving the background evolution in the kinetic equilibrium case is outlined as follows. From the above phase space distribution, we calculate the particle number density $n_{\rm{h}}$, energy density $\rho_{\rm{h}}$, pressure $p_{\rm{h}}$, and entropy density $s_{\rm{h}}$. These quantities are determined by $x$ and $\Delta$, whose values change as the universe expands. We derive the differential equations [see Eq.~\eqref{eq:differential-x-Delta}] for $x$ and $\Delta$ by requiring the comoving number density $n_{\rm{h}}a^3$ and the comoving entropy density $s_{\rm{h}}a^3$ to be conserved. After solving for the evolution of $x$ and $\Delta$, we use Eqs.\,\eqref{eq:HiDM-number-density}-\eqref{eq:HiDM-entropy-density} to calculate the evolution of $n_{\rm{h}}$, $\rho_{\rm{h}}$, $p_{\rm{h}}$ and $s_{\rm{h}}$. We also derive the adiabatic sound speed [Eq.~\eqref{eq:adiabatic-sound-speed-usual-form}], which will be used later in studying scalar perturbations. The details of the above process are given in Appendix \ref{section:background-I-equilibrium}. 

\begin{figure}[tbp]
    \centering
    \includegraphics[width=1\textwidth]{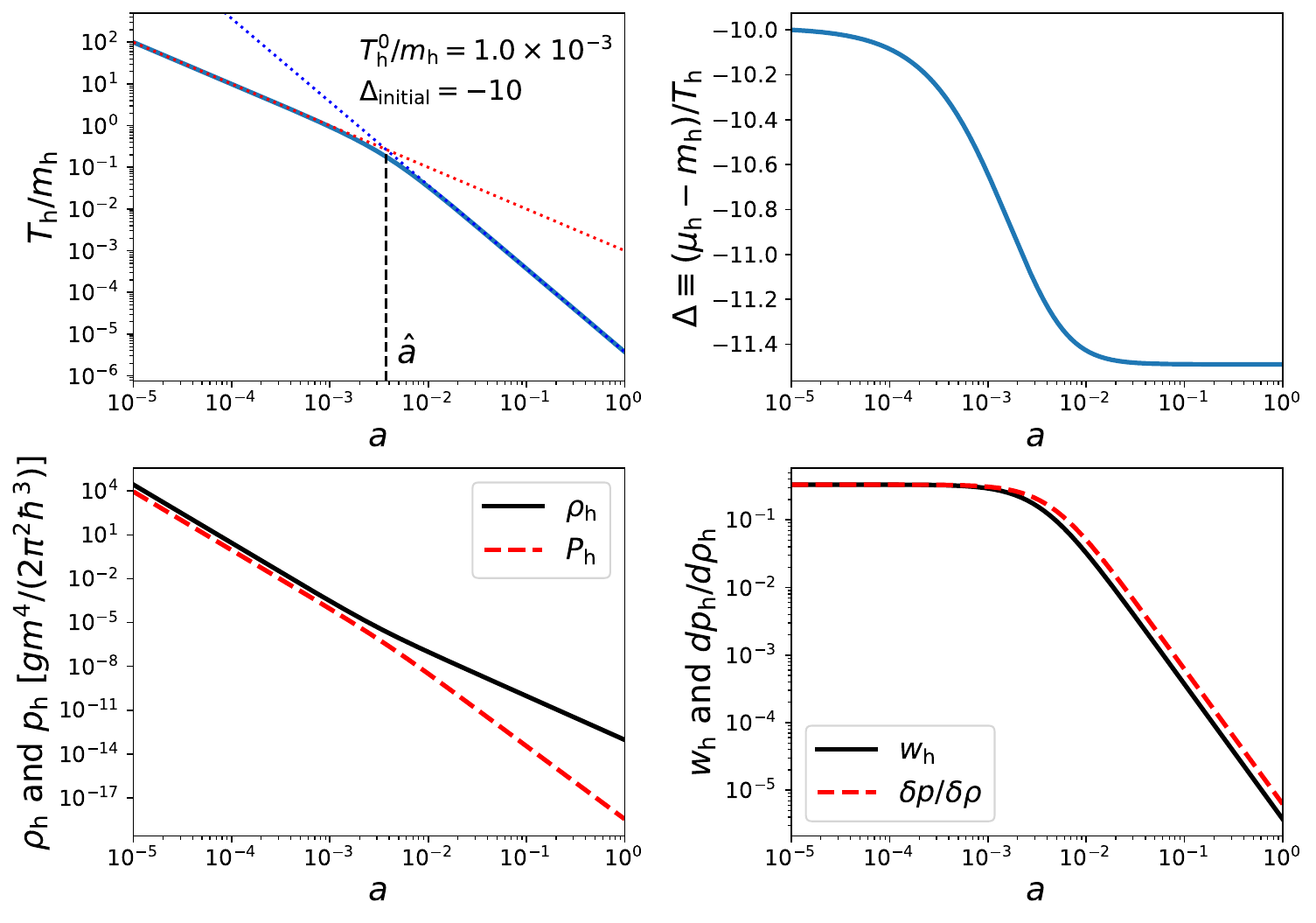}
    \caption{Example of evolution of background variables of HiDM . Here we set $T_{\rm h}^0/m_{\rm h}=10^{-3}$ and $\Delta_{\rm ini}=-10$ (a classical case). The temperature transition scale factor is indicated by $\hat{a}$. The evolution is computed by requiring the comoving particle number and specific entropy (and hence the comoving entropy density) to remain unchanged. The variables shown are the HiDM temperature-to-mass ratio $T_{\rm h}/m_{\rm h}$ (upper left), the effective chemical potential-to-temperature ratio $\Delta\equiv(\mu_{\rm h}-m_{\rm h})/T_{\rm h}$ (upper right), energy density $\rho_{\rm{h}}$ and pressure $p_{\rm{h}}$ (lower left), the equation-of-state parameter $w_{\rm{h}}\equiv p_{\rm{h}}/\rho_{\rm{h}}$ and the adiabatic sound speed $c_{\rm{s,h}}^2\equiv\delta p_{\rm{h}}/\delta\rho_{\rm{h}}$ (lower right). Different initial warmness only horizontally shift the evolution. As long as $\Delta_{\rm{ini}}\leq0$, the properties of the evolution (transition width and transition scale factor) only weakly depend on quantum statistics, which will be shown in Sec.\,\ref{section:approximation-BG-short}.}
    \label{fig:sample_result}
\end{figure}

In Figure \ref{fig:sample_result} we show an example of a solution of all the background variables as well as the adiabatic sound speed squared. In that example, we set $T_{\rm h}^0/m=10^{-3}$ and $\Delta_{\rm{ini}}=-10$ corresponding to a classical case. The evolution of the temperature is shown in the upper left panel of Figure \ref{fig:sample_result} along with its asymptotic behavior at the relativistic and nonrelativistic limits. As expected, the temperature of the HiDM (and $1/x$) drops initially as $1/a$ in the relativistic phase (when $x\rightarrow0$) and then as $1/a^2$ in the nonrelativisitc phase (when $x\rightarrow\infty$). The value of $\Delta$ (upper right panel) changes from one constant to another constant, which is shown in the upper middle panel; see Appendix \ref{section:background-I-equilibrium} for more discussion. We find that as long as $\Delta_{\rm{ini}}\leq0$, $\Delta$ is always dropping regardless of the initial conditions or whether the HiDM is a fermion or boson. This means that quantum statistical effects will be less significant after the transition. For bosonic HiDM, a Bose-Einstein condensation would never take place at the background level if $\Delta_{\rm{ini}}<0$. The drop of $\Delta$ (denoted as $\Delta_{\rm ch}$) can be calculated analytically; see Eq.~\eqref{eq:Delta-consistency}. It is always finite and approaching a constant when $\Delta_{\rm{ini}}\ll-1$. When $\Delta\gtrsim-1$, $\Delta_{\rm ch}$ only weakly depends on $\Delta_{\rm{ini}}$ and whether the HiDM is a fermion or boson. We investigate such a $\Delta_{\rm{ini}}$ dependence of $\Delta_{\rm ch}$ in Sec.\,\ref{section:dependence-on-Delta}. We have verified our numerical solutions by checking that the comoving number density and comoving entropy density are constant. Both of them are constant, as expected. The energy density and the pressure are shown in the lower left panel of Figure \ref{fig:sample_result}. The energy density first drops as $1/a^4$ (as radiation) and then as $1/a^3$ (behaving as the usual cold dark matter). The pressure first follows the energy density and drops as $1/a^4$ and then as $1/a^5$. The equation-of-state parameter $w_{\rm{h}}\equiv p_{\rm{h}}/\rho_{\rm{h}}$, as shown in the lower right panel, exhibits a smooth transition.

\subsubsection{Background evolution II: free-streaming case}\label{section:background-fs-case}
If HiDM becomes free streaming before the relativistic-to-nonrelativistic transition, its background-level physical momentum distribution at an arbitrary $a$ is given by \cite{Lewis-Challinor-2002}
\begin{equation}\label{eq:HiDM-free-stream-p-distribution-complete}
    f_{\rm{fs}}(p;\,T_{\rm h}^0, a) = \frac{g}{2\pi^2\hbar^3}\frac{p^2}{\exp\left(\frac{\sqrt{p^2a^2-m_{\rm h}^2a_{\rm fs}^2}}{T_{\rm h}^0}-\Delta_{\rm ini}\right)\pm1}\,.
\end{equation}
Here, $T_{\rm h}^0$ has the same definition as the kinetic equilibrium case, and $a_{\rm{fs}}$ is the scale factor when HiDM became free-streaming.\footnote{Note that $T_{\rm h}^0$ is not the dark matter temperature today. Particles with a momentum distribution of Eq.~\eqref{eq:HiDM-free-stream-p-distribution-complete} are not in equilibrium at the nonrelativistic phase, and $T_{\rm h}^0$ is then not the temperature of the hidden sector at the nonrelativistic phase. In fact, there should not be a temperature assigned at the nonrelativistic phase for the free-streaming case. One can define an effective temperature based on the velocity dispersions at the nonrelativistic phase, which would be very different from $T_{\rm h}^0$ }  Since we assume in this case HiDM became free-streaming at the relativisitic phase, we have $m_{\rm h}a_{\rm fs}\ll T_{\rm h}^0$ and Eq.~\eqref{eq:HiDM-free-stream-p-distribution} reduces to,
\begin{equation}\label{eq:HiDM-free-stream-p-distribution}
    f_{\rm{fs}}(p;\,T_{\rm h}^0, a) = \frac{g}{2\pi^2\hbar^3}\frac{p^2}{\exp\left(pa/T_{\rm h}^0-\Delta_{\rm ini}\right)\pm1}\,.
\end{equation}
When $\Delta_{\rm ini}=0$, the free-stream case recovers the thermal WDM scenario \cite{Colombi-Scott-Widrow-1996}. 

\begin{figure}
    \centering
    \includegraphics[width=\textwidth]{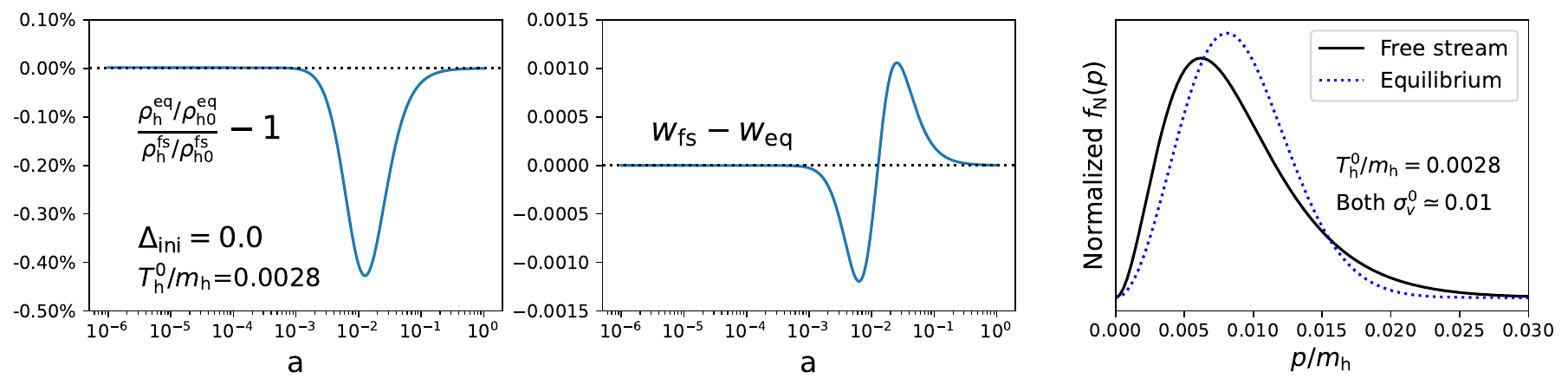}
    \caption{A comparison between the free-streaming case and the kinetic equilibrium case. In this example, we consider a fermionic HiDM with $\Delta_{\rm{ini}}=0$. In both cases, HiDM has $T_{\rm h}^0/m=0.0028$. The energy densities in the two cases (left panel) only differ from each other by less than $0.5\%$ at all times. The equation-of-state parameters (middle panel) are close to each other as well. The two nonrelativistic momentum distributions today (right panel) are different, but their velocity dispersions only differ from each other by $3\%$.}
    \label{fig:fs-eq-comparison}
\end{figure}

Given the above distribution Eq.~\eqref{eq:HiDM-free-stream-p-distribution}, the energy density and pressure at an arbitrary $a$ can be calculated. We find that the background evolution as derived from Eq.~\eqref{eq:HiDM-free-stream-p-distribution} is very similar to that of the kinetic equilibrium case. In Figure \ref{fig:fs-eq-comparison} we show a comparison of energy density, equation-of-state parameter, and the nonrelativistic momentum distribution between the free-streaming case and the kinetic equilibrium case, taking a fermionic  as an example. Starting from the same initial warmness, the energy densities (left panel) in the two cases only differ from each other by less than $0.5\%$ at all times. Also, although the momentum distributions at the nonrelativistic phase (right panel) are somewhat different, they give a similar velocity dispersion with a fractional difference of only $3\%$. 

Since the background evolution in the free-streaming case is similar to that of the kinetic equilibrium case, we will use the kinetic equilibrium case at the background level as a case study to investigate the constraints on the HiDM model. Constraints should be only slightly different between different cases. We will however distinguish them at the perturbation level; see Sec.\,\ref{section:perturbation}.

\subsection{A universal approximation to numerical solutions}\label{section:approximation-BG-short}
It is beneficial to have a good parameterized function to approximate the numerical solution of the background evolution. This can not only simplify and speed up the parameter inference process but also usually allow us to identify some properties that characterize the evolution. 
We find that the following parameterized functions can very well approximate the evolution of the energy density $\rho_{\rm{h}}$, the equation-of-state parameter $w_{\rm{h}}$, and the squared adiabatic sound speed $c_{\rm{s}}^2$ for all cases in the HiDM model:
\begin{align}
    \rho_{\rm{h}}&=\frac{\Omega_{\rm{h}}\rho_{\rm{c}}^0}{a^4}\Big[\frac{a^{n_w}+(a_{\rm{T}}^w)^{n_w}}{1+(a_{\rm{T}}^w)^{n_w}}\Big]^{\frac{1}{n_w}}\,,\label{eq:HiDM-rho-parameterization-final}\\
    w_{\rm{h}}&= \frac{1}{3}\frac{(a_{\rm{T}}^w)^{n_w}}{a^{n_w}+(a_{\rm{T}}^w)^{n_w}}\,,\label{eq:HiDM-w-parameterization-final}\\
    c_{\rm{s}}^2&=\frac{1}{3}\frac{(a_{\rm{T}}^{c_{\rm{s}}^2})^{n_{c_{\rm{s}}^2}}}{a^{n_{c_{\rm{s}}^2}}+(a_{\rm{T}}^{c_{\rm{s}}^2})^{n_{c_{\rm{s}}^2}}}\,,\label{eq:HiDM-cs2-parameterization-final}
\end{align}
where $\rho_{\rm{c}}$ is today's critical energy density. The transition indices $n_w$ and $n_{c_{\rm{s}}^2}$ quantify the width in $\ln a$ of the relativistic-to-nonrelativistic transitions of $w_{\rm{h}}$ and $c_{\rm{s}}^2$, respectively. The larger the transition index, the sharper the transition. Both $w_{\rm{h}}$ and $c_{\rm{s}}^2$ start with $1/3$ and transition to $0$. The scale factors $a_{\rm{T}}^w$ and $a_{\rm{T}}^{c_{\rm{s}}^2}$ roughly represent the position in $a$ where $w_{\rm{h}}$ and $c_{\rm{s}}^2$ begin to drop. Note that the functional form of $\rho_{\rm{h}}$ is consistently obtained from the above parameterized $w_{\rm{h}}$ according to $d(\rho_{\rm{h}}a^3)=-w_{\rm{h}}\rho_{\rm{h}}d(a^3)$. But, for a better approximation to the numerical result of $c_{\rm{s}}^2$, we use a separate parameterized function [Eq.~\eqref{eq:HiDM-cs2-parameterization-final}] to fit the evolution of $c_{\rm{s}}^2$, rather than calculating it using $dp_{\rm h}/d\rho_{\rm h}$.

Both $a_{\rm{T}}^w$ and $a_{\rm{T}}^{c_{\rm{s}}^2}$ are proportional to $T_{\rm h}^0/m$. We define two proportionality coefficients as
\begin{equation}\label{eq:alpha-T0-m}
    \alpha_{w} = \frac{a_{\rm{T}}^{w}}{T_{\rm h}^0/m}\,,
    ~~~~ \textrm{and} ~~~~~
    \alpha_{c_{\rm{s}}^2} = \frac{a_{\rm{T}}^{c_{\rm{s}}^2}}{T_{\rm h}^0/m}\,.
\end{equation}
There are small differences between the scale factors $a_{\rm{T}}^w$ and $a_{\rm{T}}^{c_{\rm{s}}^2}$. We shall show that the differences among these three scale factors are independent of the initial warmness and vary only slightly for different cases. To isolate the dependence on the initial warmness, we quantify the difference between the two scale factors on a $\log_{10}$ scale and define 
\begin{equation}
    \delta_a\equiv\log_{10}a_{\rm{T}}^{c_{\rm{s}}^2}-\log_{10}a_{\rm{T}}^w\,,\label{eq:delta_w_definition}
\end{equation}
Then, given $a^w_{\rm T}$, the three transition parameters, $n_w$, $n_{c_{\rm{s}}^2}$ and $\delta_a$ determine the background and sound speed evolution as parameterized by Eqs.\,\eqref{eq:HiDM-rho-parameterization-final}-\eqref{eq:HiDM-cs2-parameterization-final}. We summarize the meanings of some parameters in Table \ref{tab:summary_of_variables}.

\begin{figure}[tbp]
    \centering
    \includegraphics[width=\textwidth]{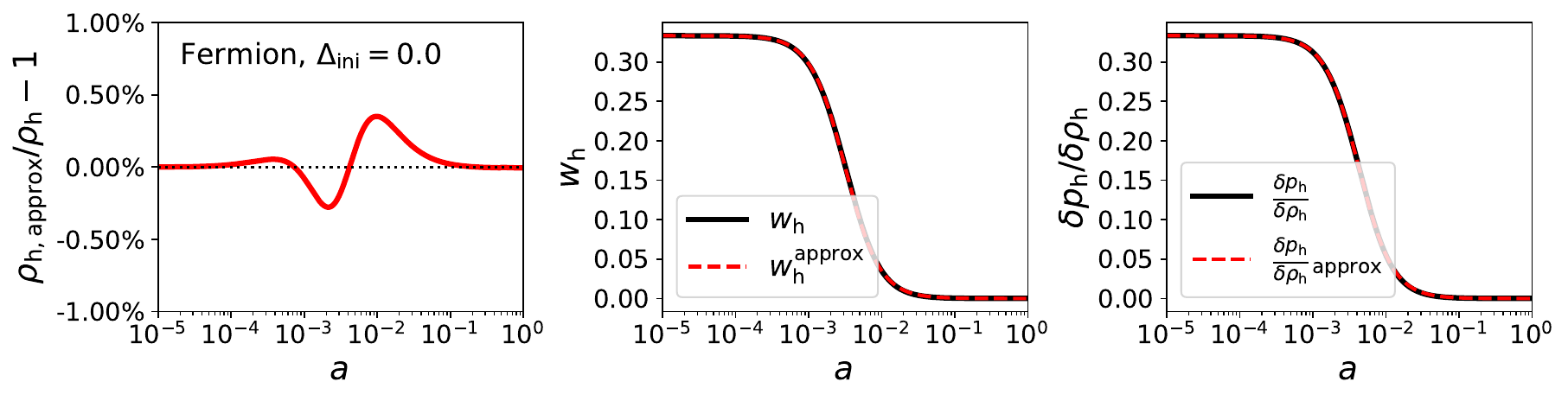}
    \includegraphics[width=\textwidth]{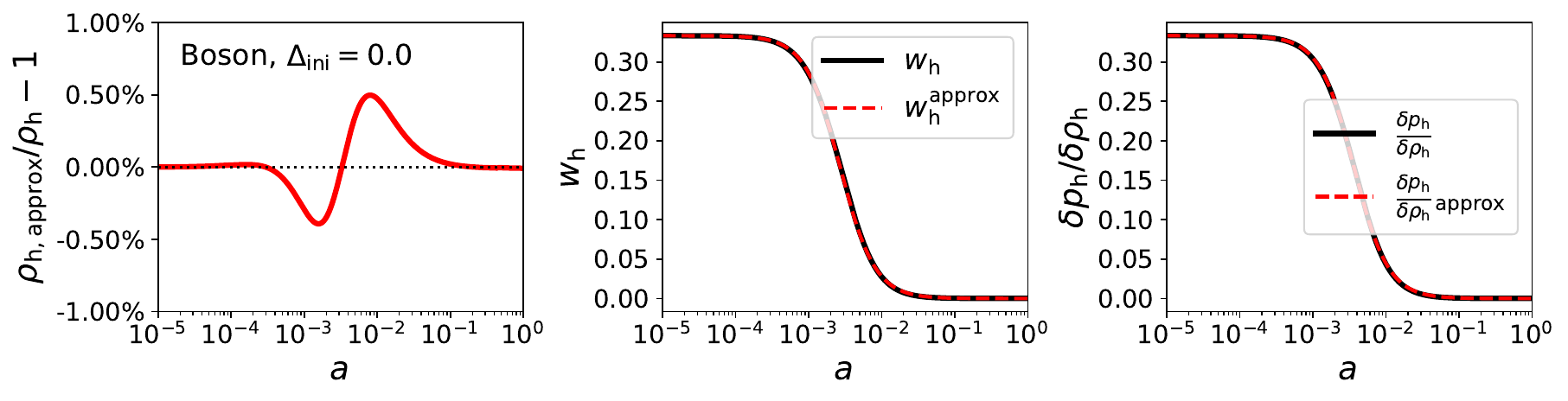}
    \includegraphics[width=\textwidth]{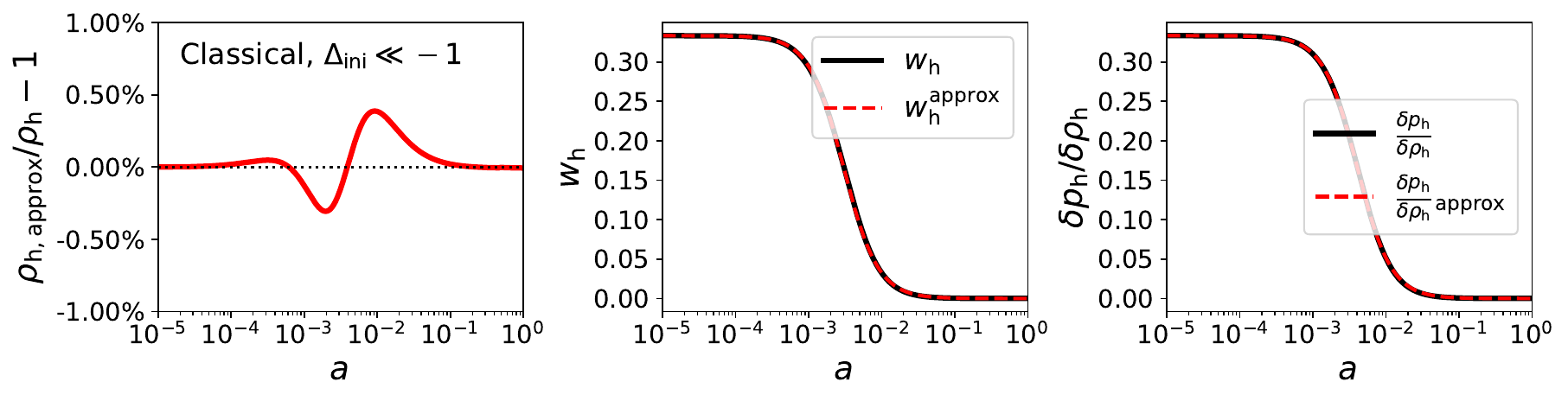}
    \caption{Comparison between the numerical solutions and approximations for a fermion case with $\Delta_{\rm{ini}}=0$ (first row), a boson case with $\Delta_{\rm{ini}}=0$ (secon row), and the classical case where $\Delta{\rm ini}\ll-1$ (third row, we take $\Delta_{\rm ini}=-10$ here). Left column: the fractional difference between the numerical solution of $\rho_{\rm{h}}$ and its best-fit approximation. The fitting function is Eq.~\eqref{eq:HiDM-rho-parameterization-final}. The fractional difference is within $0.5\%$ at all times for all cases considered here. Middle column: the numerical solution of $w_{\rm{h}}$ (solid black) and its approximation [dashed red, calculated with Eq.~\eqref{eq:HiDM-w-parameterization-final}]. Right column: $\delta p_{\rm{h}}/\delta \rho_{\rm{h}}$ and its approximation. The numerical solution and the approximation match very well in all cases.}
    \label{fig:approx-w-dpdrho}
\end{figure}

We fit the parameterized functions to the numerical solutions of $\rho_{\rm{h}}$ and $c_{\rm{s}}^2$ to find the best-fit transition parameters. The fitting process is detailed in Appendices \ref{section:approximation-I-equilibrium} and \ref{section:approximation-fs}. We show the best-fit transition parameters along with other useful parameters in Table \ref{tab:transition-parameters}. Figure \ref{fig:approx-w-dpdrho} presents examples comparing the fitting functions with numerical solutions for HiDM in kinetic equilibrium. We include three cases: a fermionic scenario with $\Delta_{\rm{ini}}=0$, a bosonic scenario with $\Delta_{\rm{ini}}=0$, and a classical case where $\Delta_{\rm{ini}} \ll -1$ (illustrated with $\Delta_{\rm{ini}}= -10$ in this example). The fractional difference in the energy density between the approximation and the numerical solution is within about $0.5\%$ at all times. The approximations to the equation-of-state parameter and the adiabatic sound speed are also very good. 

For the free-streaming case, we use a similar procedure to fit the same parameterized function Eqs.\,\eqref{eq:HiDM-rho-parameterization-final} to the numerical solution of $\rho_{\rm{h}}$. We include an example of the fitting for the free-streaming case in Appendix \ref{section:approximation-fs}. Note that a sound speed is not applicable for the free-streaming case. 

Different cases have the very similar behavior in the relativistic-to-nonrelativistic transition, implying that the observational effects are similar for all cases we consider. Indeed, for the background evolution, we can see from our approximations [Eq.~\eqref{eq:HiDM-rho-parameterization-final}-\eqref{eq:HiDM-cs2-parameterization-final}] that the important properties are the transition position, the transition width and the transition gap between $w_{\rm{h}}$ and $c_{\rm{s}}^2$. Importantly, the latter two are very similar for all cases [see the rows of $n_w$, $n_{c_{\rm{s}}^2}$ and $\delta_a$ in Table \ref{tab:transition-parameters}]. The transition positions in different cases can be brought to agreement by changing the initial warmness. These similarities show that it is justified to take the background evolution in the classical case as our representative case to study the impacts on cosmological observables and their associated constraints. Thus, the most important parameter is the transition scale factor.

The background evolution in our representative case (the classical case) only depends the transition scale factor $a_{\rm T}^w$. We define the following parameter that represents how early the transition compared to matter-radiation equality,
\begin{equation}\label{eq:a_t-parameterization}
    R^{\rm{T}}_{\rm{eq}} \equiv \frac{a_{\rm T}^w}{a_{\rm{eq}}}\,,
\end{equation}
where $a_{\rm{eq}} = \frac{4.15\times10^{-5}}{\Omega_{\rm{m}}h^2}$ is the scale factor at matter-radiation equality in the standard CDM case. With this definition, $R^{\rm{T}}_{\rm{eq}}<1$ means that the transition happened before the standard matter-radiation equality. The standard cold dark matter case corresponds to $R^{\rm{T}}_{\rm{eq}}\rightarrow0$. 

Now, the evolution of HiDM with conservation of comoving number density and comoving entropy can be well parameterized by only two parameters: $\Omega_{\rm{h}}$ and $R^{\rm{T}}_{\rm{eq}}$. Therefore, we have only one additional parameter compared to the standard $\Lambda$CDM model: $R_{\rm{eq}}^{\rm{T}}$. This one-parameter family of dark matter models is similar to that of WDM proposed in Ref.\,\cite{Colombi-Scott-Widrow-1996} but ours is more generalized, allowing different quantum statistics, initial chemical potential, and dark matter models that are either in their own kinetic equilibrium or free streaming. The conclusions about the impacts on cosmological observations and parameter estimations presented later in this work are independent of the mass or the intrinsic degrees of freedom. Additionally, they are fairly insensitive to whether the dark matter is free streaming or in kinetic equilibrium, and whether it is composed of fermions or bosons (as long as $\Delta_{\rm ini}\leq0$).

\begin{table}[]
    \centering
    \begin{tabular}{c|l}
    \hline
    \hline
    Parameters    & Meanings  \\
    \hline
    $\Delta$      & Effective chemical potential-to-temperature ratio. \\
    $x$           & Mass-to-temperature ratio.\\
    $T_{\rm h}^0/m$ & Initial warmness $\equiv\lim_{a\rightarrow0}aT_{\rm h}/m_{\rm h}$.\\
    $a_{\rm T}^w$ ($a_{\rm T}^{c_{\rm s}^2}$) & Scale factor at which the DM equation of state (sound speed) drops.\\
    $\alpha_w$ ($\alpha_{c_{\rm s}^2}$)    & Proportional coefficient between $a_{\rm T}^w$ ($a_{\rm T}^{c_{\rm s}^2}$) and the initial warmness.\\
    $n_w$ ($n_{\rm c_{\rm s}^2}$) & The transition index of DM equation of state (sound speed).\\
    $\delta_a$ & The gap between $a_{\rm T}^w$ and$a_{\rm T}^{c_{\rm s}^2}$ in the logarithmic scale. \\
    $\hat{a}$     & Transition scale factor at which the DM temperature drops.\\
    $C$           & The proportional coefficient between $\hat{a}$ and the initial warmness.\\ 
    \hline
    \end{tabular}
    \caption{Summary of parameters and their meanings.}
    \label{tab:summary_of_variables}
\end{table}

\begin{table}[tbp]
    \centering
    \begin{tabular}{lrrr|rrr}
    \hline
    \hline
            & \multicolumn{3}{c|}{Kinetic equilibrium} & \multicolumn{3}{c}{Free-Streaming}\\
            & \multicolumn{1}{c}{$\Delta_{\rm{ini}}\ll-1$} & \multicolumn{1}{c}{$\Delta_{\rm{ini}}=0$, F} & \multicolumn{1}{c|}{$\Delta_{\rm{ini}}=0$, B~~} & \multicolumn{1}{c}{$\Delta_{\rm{ini}}\ll-1$} & \multicolumn{1}{c}{$\Delta_{\rm{ini}}=0$, F} &\multicolumn{1}{c}{$\Delta_{\rm{ini}}=0$, B} \\
            \hline
        $\alpha_w$ & 3.0002 & 3.1529 & 2.6942 & 3.0003 & 3.1515  & 2.7013\\
        $\alpha_{c_{\rm{s}}^2}$ & 4.0208 & 4.2217 & 3.6214 & N/A & N/A & N/A \\
        $n_{w}$  & 1.8442 & 1.8539 & 1.8197 & 1.8169 & 1.8367 & 1.7838 \\
        $n_{c_{\rm{s}}^2}$ & 1.8707 & 1.8820 & 1.8463 & N/A & N/A & N/A \\
        $\delta_a$ & 0.1275 & 0.1268 & 0.1288 & N/A & N/A & N/A\\
        $C$ & 3.7120& 3.9203 & 3.2802 & N/A & N/A & N/A \\
        $\Delta_{\rm ch}$ & $-1.5$ & -1.6202 & -1.2451 & N/A & N/A & N/A\\
        \hline
    \end{tabular}
    \caption{Summary of the transition parameters. The coefficient $\alpha_{(w,c_{\rm{s}}^2)}$ is defined in Eq.~\eqref{eq:alpha-T0-m} and $C$ is defined via Eqs.\,\eqref{eq:x-relativistic-limit} and \eqref{eq:x-nonrelativisitic-limit}. The change of $\Delta$ for the kinetic equilibrium case is defined as $\Delta_{\rm ch}\equiv\Delta_{\rm{f}}-\Delta_{\rm{ini}}$. The parameters $n_w$, and $n_{c_{\rm{s}}^2}$ are the best-fit transition indices of the approximations to $\rho_{\rm{h}}$ and $c_{\rm{s}}^2$. The fitting functions are Eq.~\eqref{eq:HiDM-rho-parameterization-final} for $\rho_{\rm{h}}$ and Eq.~\eqref{eq:HiDM-cs2-parameterization-final} for $\delta p_{\rm{h}}/\delta \rho_{\rm{h}}$. The gap $\delta_a\equiv\log_{10}a_{\rm{T}}^{w}-\log_{10}a_{\rm T}^{c_{\rm{s}}^2}$. In the kinetic equilibrium case, the classical limits ($\Delta_{\rm ini}\ll-1$) are given in the second column. A fermion (boson) case with $\Delta_{\rm{ini}}=0$ is shown in the third (fourth) column. ``F'' and ``B'' stand for fermion and boson. The corresponding free-streaming cases are shown in the last three columns. \newline In the rest of this work we will use the transition parameters at the classical limit to study impacts on cosmological observations and parameter estimations; see Appendix \ref{section:dependence-on-Delta} for more discussions about the phenomenologically weak dependence on different cases. Values of $C$ and $\Delta_{\rm ch}$ are derived analytically [see Eq.~\eqref{eq:Delta-consistency} and Eq.~\eqref{eq:C-consistency}] and verified by numerical results. In particular, the value of $C=3.712$ in classical limit in the kinetic equilibrium case corresponds to $\big(\frac{8}{\pi}\big)^{\frac{1}{3}}\exp(1)$; see the discussion in Sec.\,\ref{section:approximation-I-equilibrium}.}
    \label{tab:transition-parameters}
\end{table}

\subsection{Remarks on the transition fitting formula}\label{sec:remarks_on_background}

We note that the motivation to use these prameterizatized functions (Eqs.\,\eqref{eq:HiDM-rho-parameterization-final} to \eqref{eq:HiDM-cs2-parameterization-final}) is to approximate the evolution of those quantities. The evolution of the density is sometime more physically related to cosmological observables than the equation of state. Thus, we directly use the parameterized density evolution Eq.~\eqref{eq:HiDM-w-parameterization-final} to fit the numerical result, instead of the parameterized equation of state. Another advantage of our parameterization is that an analytic form of $\rho_{\rm{h}}(a)$ can be obtained, which expedites numerical calculations. However, particular attention needs to be paid to Eq.~\eqref{eq:HiDM-w-parameterization-final}. This is because Eq.~\eqref{eq:HiDM-w-parameterization-final} does not represent a correct late-time evolution of the equation of state, which would drop as $\propto1/a^2$ instead of $\propto1/a^{n_w}$. But, $n_w\simeq1.8$ is close to $2$. An important quantity is the HiDM sound horizon (or the free-streaming scale for the free-streaming case) that determines the suppression scale of the matter power spectrum. We calculate the sound horizon respectively with the numerical and the approximated function and find that they only differ by less than $3\%$.

Other forms of parameterization can be found in the literature. For example, Ref.\,\cite{Hu1998ApJ,Ganjoo:2022rhk} uses 
\begin{equation}\label{eq:fitting-Hu1998}
    w(a)=\frac{1}{3}(1+(a/a_{\rm{p}})^{2p})^{-\frac{1}{p}}\,.
\end{equation}
This fitting formula can correctly represent a late-time behavior but couldn't be integrated to give an analytical density expression. For a completeness, we also provide the best-fit parameters using this fitting formula for the HiDM equation of state and adiabatic sound speed, which we show in Table \ref{tab:transition-parameters-H1998}.

Another form adopted in Ref.\,\cite{2018Das-etal-Ballistic-DM}, $w=\frac{1}{6}\big[1-\tanh(\frac{a-a_*}{\Delta})\big]$,  does not work in our case. This is because, under this parameterization, $w$ would either exhibit a very sharp transition (e.g., when $\Delta/a_*=0.1$) or not start with $1/3$ (e.g., when $\Delta/a_*=1$).

\begin{table}[tbp]  
    \centering
    \begin{tabular}{crrr|rrr}
    \hline
    \hline
            & \multicolumn{3}{c|}{Kinetic equilibrium} & \multicolumn{3}{c}{Free-Streaming}\\
            & \multicolumn{1}{c}{$\Delta_{\rm{ini}}\ll-1$} & \multicolumn{1}{c}{$\Delta_{\rm{ini}}=0$, F} & \multicolumn{1}{c|}{$\Delta_{\rm{ini}}=0$, B~~} & \multicolumn{1}{c}{$\Delta_{\rm{ini}}\ll-1$} & \multicolumn{1}{c}{$\Delta_{\rm{ini}}=0$, F} &\multicolumn{1}{c}{$\Delta_{\rm{ini}}=0$, B} \\
            \hline
        $\frac{a_{{\rm p},w}}{(T_{\rm h}^0/m)}$  & 3.3691  & 3.5124  & 3.0873  & 3.4125  & 3.5442  & 3.1774  \\
        $p_w$ & 0.8768  & 0.8837  & 0.8606  & 0.8623  &  0.8723  & 0.8353  \\
        $\frac{a_{{\rm p},{c_{\rm s}^2}}}{(T_{\rm h}^0/m)}$  & 4.3560  & 4.5385  & 3.9946  & N/A  & N/A  & N/A  \\
        $p_{c_{\rm s}^2}$ & 0.9062  & 0.9142  & 0.8881  & N/A  & 0.N/A  & 0.N/A  \\
        \hline
    \end{tabular}
    \caption{\label{tab:transition-parameters-H1998}Best-fit parameters using the fitting function Eq.~\eqref{eq:fitting-Hu1998} to HiDM equation of state and the adiabatic sound speed.}
\end{table}

\subsection{The asymptotic behaviors of the mass-to-temperature ratio and the mass density today}

For the kinetic equilibrium case, the mass-to-temperature ratio $x$ goes as $a$ at early times when $x\ll1$ and as $a^2$ at late times when $x\gg1$. Such asymptotically relativistic and nonrelativistic behaviors are shown by the dotted lines in the upper left panel of Figure \ref{fig:sample_result} where the reciprocal -- the temperature-to-mass ratio $1/x$ -- is plotted. We define $\hat{a}$ as the scale factor where the two asymptotic lines meet so that the two limits can be written as
\begin{align}
    \lim_{x\rightarrow0}x&=C\frac{a}{\hat{a}}\,, \label{eq:x-relativistic-limit}\\
    \lim_{x\rightarrow\infty}x&=C\frac{a^2}{\hat{a}^2}\,\label{eq:x-nonrelativisitic-limit} \,,
\end{align}
Equivalently, Eq.~\eqref{eq:x-relativistic-limit} gives $\hat{a}=C\,T_{\rm h}^0/m_{\rm h}$, that is, given the initial warmness we can get $\hat{a}$. By substituting Eqs.\,\eqref{eq:x-relativistic-limit} and \eqref{eq:x-nonrelativisitic-limit} into the relativistic and nonrelativistic forms of the number density, we can derived $C$ analytically which reads,
\begin{equation}\label{eq:C-consistency}
    C=\frac{2}{\pi^{1/3}}\Big[\frac{f_3^\mp(\Delta_{\rm{ini}})}{f_{3/2}^\mp(\Delta_{\rm{f}})}\Big]^{\frac{2}{3}}\exp\big(-\frac{2}{3}\Delta_{\rm ch}\big)\,,
\end{equation}
where $\Delta_{\rm{f}}$ is the final value of $\Delta$, $\Delta_{\rm ch}\equiv\Delta_{\rm{f}}-\Delta_{\rm{ini}}$ is the change of $\Delta$ during the relativistic-to-nonrelativistic transition and is given analytically by Eq.~\eqref{eq:Delta-consistency}, and the function $f_s^{\mp}$ defined in Eq.~\eqref{eq:Fs} is of the order unity.

In the usual WDM scenario, the relativisitic-to-nonrelativistic transition time (parameterized by $R_{\rm{eq}}^{\rm{T}}$) can be determined from by the dark matter density fraction today and the particle mass \cite{Colombi-Scott-Widrow-1996,Viel-etal-2005}. In our HiDM scenario, the relation between the particle mass and the transition time is modified and given by (see Eq. (16) in \cite{Zhang:2024vkx} with a slight modification)
\begin{equation}\label{eq:Om-Req-chi}
\Omega_{\rm dm} h^2 = 0.12\times\frac{g}{2}\,\gamma(\Delta_{\rm ini})\Big(\frac{m}{0.86\,{\rm keV}}\Big)^4\Big(\frac{R_{\rm{eq}}^{\rm{T}}}{5.3\times10^{-4}}\Big)^3\,,
\end{equation}
where 
\begin{equation}\label{eq:gamma-factor}
    \gamma(\Delta_{\rm ini})= \left(\frac{f_{3/2}^\mp(\Delta_{\rm{f}})}{f^-_{3/2}(\Delta_{\rm{f}}^0)}\right)^{5/2}\left(\frac{f_{5/2}^\mp(\Delta_{\rm{f}}^0)}{f^\mp_{5/2}(\Delta_{\rm{f}})}\right)^{3/2}\frac{\exp(\Delta_{\rm{f}})}{\exp(\Delta_{\rm{f}}^0)}\,,
\end{equation}
where $\Delta_{\rm{f}}^0$ is the final value of $\Delta$ when $\Delta_{\rm ini}=0$. Note that in the above it is the final value $\Delta_{\rm{f}}$ that enters the $\gamma$ factor. Since $\gamma(\Delta_{\rm ini}\leq0)\geq1$, for a given mass of dark matter, a later transition (larger $R_{\rm{eq}}^{\rm{T}}$) is required for a lower value of $\Delta_{\rm{ini}}$ to account for the observed dark matter abundance. As we shall see, observations put upper bounds on $R^{\rm{T}}_{\rm{eq}}$. In the WDM scenario with $\Delta_{\rm ini}=0$, an upper bound on $R^{\rm{T}}_{\rm{eq}}$ corresponds to a lower bound on the particle mass. In the HiDM scenario, the corresponding lower bounds on the particle mass become even stronger when $\Delta_{\rm ini}<0$ since $\gamma(\Delta_{\rm ini}\leq0)\geq1$.

\subsection{Constraints from Big Bang Nucleosynthesis element abundance}
If dark matter were relativistic during BBN, the presence of extra relativistic degrees of freedom would have caused the expansion of the universe to be faster, altering the primordial element abundance. Abundance observations of, e.g., primordial deuterium and helium have independently put a constraint on the extra radiation energy \cite{2018-Cooke-etal-DH,2014-Izotov-Thuan-Yp}. This puts a constraint on the transition time for HiDM and thus an upper bound on $R^{\rm{T}}_{\rm{eq}}$. 

In Appendix \ref{section:constraint-BBN-detail} we show that  $R^{\rm{T}}_{\rm{eq}}$ is related to $\Delta N_{\rm{eff}}$ (that parameterizes the extra radiation energy density at BBN) by
\begin{equation}\label{eq:RTeq-Neff}
    R_{\rm{eq}}^{\rm{T}}=0.23\times\frac{\Omega_{\rm{m}}\Omega_{\gamma}}{\Omega_{\rm{dm}}\Omega_{\rm{rad}}}\Delta N_{\rm{eff}}\,,
\end{equation}
where $\Omega_{\rm{m}}$, $\Omega_{\rm{dm}}$, $\Omega_{\gamma}$ and $\Omega_{\rm{rad}}$ are today's density parameters of matter, dark matter, photons and total radiation.\footnote{Note that $\Omega_{\rm{rad}}$ here is calculated by treating neutrinos as relativistic particles.} By fixing all $\Omega$'s to the mean values given in Planck 2018 \cite{Planck2018-parameter-constraints} and taking the upper bound $\Delta N_{\rm{eff}}<0.4$ \cite{Yeh:2022heq}, we have the upper bound of $R_{\rm{eq}}^{\rm{T}}$ constrained by BBN,
\begin{equation}\label{eq:BBN-RTeq-upper-bound}
    R_{\rm{eq}}^{\rm{T}}\lesssim0.065\,.
\end{equation}
We have ignored the uncertainties of all $\Omega$'s, as they are sub-dominant. Compared to the other constraints that we present later, this BBN constraint is the weakest. 

We however emphasize that our $\Lambda$HiDM model is \emph{not} equivalent to an $\Delta N_{\rm{eff}}$-$\Lambda$CDM model when other cosmological and astrophysical observations are concerned. The impacts in different observations will be analyzed individually. As we shall see, other cosmological and astrophysical constraints on $R_{\rm{eq}}^{\rm{T}}$ are much stronger than the constraint from BBN; we nonetheless provide a rough estimate of the BBN constraint here.

\section{Scalar perturbations of the hidden dark matter}\label{section:perturbation}
HiDM exhibits intriguing phenomenology at the perturbation level. HiDM oscillates in the relativisitic phase, its overdensity would oscillate. If the transition from relativistic to nonrelativistic occurs sufficiently late, these oscillations can lead to a detectable suppression of small-scale structures.

In this section, we first outline the perturbation equations and the initial conditions for HiDM. Our notation primarily follows \cite{Ma-Bertschinger1995}, and we adopt the synchronous gauge. Throughout this work, we focus on scalar-mode perturbations with adiabatic initial conditions and zero spatial curvature.

\subsection{Perturbations in standard CDM}\label{section:perturbations-std}
In the standard CDM scenario, the velocity potential of dark matter is constant. This allows a gauge transformation that eliminates the velocity potential and fixes the residual degree of freedom in the synchronous gauge \cite[Ch.~5]{S.W.cosmo.}. When the velocity potential vanishes, the cold dark matter fluid dynamics are governed by
\begin{equation}\label{eq:CDM-density-differential}
    \delta_{\rm{c}}'=-h_{\rm{s}}'/2\,,
\end{equation}
where $'$ denotes differentiation with respect to the conformal time $\eta$, and $h_{\rm{s}}$ is the trace of the spatial metric perturbation in the synchronous gauge, as defined in \cite{Ma-Bertschinger1995}.

The evolution of the other synchronous gauge variable, $\eta_{\rm{s}}$, which represents the traceless and divergenceless part of the spatial metric perturbation \cite{Ma-Bertschinger1995}, is given by  
\begin{equation}\label{eq:eta_s-differential}
    k\eta_{\rm{s}}' = 4\pi Ga^2\sum\rho_i q_i\,,
\end{equation}
where $k$ is the comoving wavenumber, $\rho_i$ is the energy density of the $i$th species, and $q_i$ is its corresponding heat flux\footnote{The heat flux is related to the fluid velocity divergence $\theta$ defined in \cite{Ma-Bertschinger1995} and the velocity potential $\delta u$ defined in \cite{S.W.cosmo.} by $q=(1+w)\theta/k=-(1+w)k\delta u/a$.}. For cold dark matter, $q_{\rm{c}} = 0$.

The adiabatic initial condition for $\delta_{\rm{c}}$ to next-to-leading order is \cite{Ma-Bertschinger1995,Bucher-etal-2000-perturbation-initial-condition}
\begin{equation}\label{eq:delta_CDM_adiabatic_init}
    \delta_{\rm{c}}(k\eta\ll1) = \mathcal{R}_k\Big[-\frac{1}{4}k^2\eta^2+\frac{1}{20}\frac{\Omega_{\rm{m}}\mathcal{H}_0}{\sqrt{\Omega_{\rm{R}}}}k^2\eta^3\Big]\,,
\end{equation}
where $\delta_{\rm{c}}\equiv\delta\rho_{\rm{c}}/\bar{\rho}_{\rm{c}}$, $\mathcal{R}_k$ is the primordial curvature perturbation, $\mathcal{H}_0$ is the conformal Hubble parameter today, and $\Omega_{\rm{m}}$ and $\Omega_{\rm{R}}$ are the present-day matter and radiation density fractions, respectively.

\subsection{Perturbations in hidden dark matter}\label{section:perturbation-hb}
The primary modification to the perturbation equations in the HiDM scenario arises from the fluid dynamics of HiDM, where the bulk velocity of HiDM must be explicitly accounted for.

For the kinetic equilibrium case, the HiDM fluid dynamics follow \cite{Ma-Bertschinger1995}:
\begin{align}
    \delta_{\rm{h}}'&=-kq_{\rm{h}}-(1+w_{\rm{h}})h_{\rm{s}}'/2+3\mathcal{H}(w_{\rm{h}}-\frac{\delta p_{\rm{h}}}{\delta \rho_{\rm{h}}})\delta_{\rm{h}}\,,\label{eq:HiDM-delta-dot}\\
    q_{\rm{h}}'&=(3w_{\rm{h}}-1)\mathcal{H}q_{\rm{h}}+k\frac{\delta p_{\rm{h}}}{\delta \rho_{\rm{h}}}\delta_{\rm{h}}\,.\label{eq:HiDM-velocity-dot}
\end{align}
Here, $q_{\rm{h}}$ represents the heat flux, and $w_{\rm{h}}$ is the HiDM equation-of-state parameter. Due to the nonzero $q_{\rm{h}}$, an additional source term appears in the evolution equation for $\eta_{\rm{s}}$:
\begin{equation}\label{eq:additional-term-in-hs}
     k\eta_{\rm{s}}' = 4\pi Ga^2\Big(\sum\rho_iq_i+\rho_{\rm{h}}q_{\rm{h}}\Big)\,,
\end{equation}
where $i$ goes through all components except for HiDM.

We assume that perturbations in the HiDM scenario are seeded by adiabatic initial conditions. Note that dark matter isocurvature modes can naturally arise when the dark matter does not begin in thermal contact with the Standard Model sector; see e.g. Ref.\,\cite{Tenkanen-2019-nonthermally-produced-DM}. However, the primordial isocurvature component has been tightly constrained by Planck \cite{Planck2018inflation}, so we will assume adiabatic initial conditions. Since HiDM is initially relativistic, its adiabatic initial conditions differ from those of standard CDM and instead resemble those of photons \cite{Ma-Bertschinger1995,Bucher-etal-2000-perturbation-initial-condition}. Keeping next-to-leading order terms,  
\begin{equation}\label{eq:HiDM-initial-adiabatic}
    \delta_{\rm{h}}(k\eta\ll1)=\delta_{\gamma}(k\eta\ll1)=\mathcal{R}_k\Big[-\frac{1}{3}k^2\eta^2+\frac{1}{15}\frac{\Omega_{\rm{b}}\mathcal{H}_0k^2\eta^3}{\sqrt{\Omega_{\rm{R}}+a_{\rm T}^w\,\Omega_{\rm{h}}}}\Big]\,.
\end{equation}
Note the coefficient of the next-to-leading order term above is now different from that in Eq.~\eqref{eq:delta_CDM_adiabatic_init}, i.e., it is $\Omega_{\rm b}$ in the numerator now, because the HiDM in this case initially behaves as a type of radiation. The initial conditions of $q_{\rm{h}}$ also equal those of the photon heat flux \cite{Ma-Bertschinger1995,Bucher-etal-2000-perturbation-initial-condition}, i.e.,
\begin{equation}\label{eq:HiDM-heat-flux-initial}
    q_{\rm{h}}(k\eta\ll1)=-\frac{\mathcal{R}_k}{27}k^3\eta^3\,.
\end{equation}

For the kinetic equilibrium case, we implement the background and perturbation equations by modifying \textsc{camb} \cite{Lewis-Challinor-Lasenby-2000}. More specifically, at the background level, we modify the energy density of DM according to the fitting formula of Eq.~\eqref{eq:HiDM-rho-parameterization-final}. The fluid dynamics of DM is changed according to Eqs.~\eqref{eq:HiDM-delta-dot}, \eqref{eq:HiDM-velocity-dot} and \eqref{eq:additional-term-in-hs}. The adiabatic initial conditions of HiDM and photons are modified according to Eq.~\eqref{eq:HiDM-initial-adiabatic}. 

Additionally, there is a subtlety in defining the HiDM density. Unlike standard CDM, here the equation-of-state parameter $w_{\rm{h}}$ is not negligible, and the velocity potential is not constant. This prevents the use of a simple gauge transformation to eliminate the velocity potential. Fortunately, if the transition from relativistic to nonrelativistic occurs well before last scattering (as required in viable models), the velocity potential can be considered constant long before last scattering. Consequently, the HiDM bulk velocity (and thus the heat flux) decays to a negligible level. This is evident from the invariant density fluctuation in the synchronous gauge \cite{S.W.cosmo.},  
\begin{equation}\label{eq:synchrongauge-invariant-density}
    \tilde{\delta}_{\rm h}=\delta_{\rm h}+3(1+w_{\rm h})\frac{1}{k}\frac{a'}{a^2}q_{\rm h}\,,
\end{equation}
where the second term becomes insignificant at late times.

For the free-streaming case, the behavior closely resembles that of the WDM model, with the key difference being the nonzero $\Delta_{\rm ini}$ in Eq.~\eqref{eq:HiDM-free-stream-p-distribution}. To model a free-streaming HiDM scenario with an arbitrary phase-space distribution, we use \textsc{class} \cite{2011JCAP...07..034B} to specify a customized distribution function. The input parameters for this function are $\Delta_{\rm ini}$ and the dark matter temperature-to-photon temperature ratio, which we infer from the value of $R^{\rm T}_{\rm eq}$.

We then compare the HiDM results for both the kinetic equilibrium and free-streaming cases (each with the same $R_{\rm eq}^{\rm T}$) against the standard CDM case. Since we treat the kinetic equilibrium and free-streaming scenarios separately, we ensure consistency by setting identical parameters for the standard CDM case in both \textsc{camb} and \textsc{class}. We verify this consistency by confirming that the matter power spectra generated by the two codes are identical for the standard CDM scenario.

\subsection{Impacts on the transfer functions}
The effects of HiDM at the perturbation level are described by the modified transfer functions, which characterize the growth of perturbations at a given time and scale with respect to their initial conditions. 
Like WDM, the primary effect of HiDM is a suppression of the matter power spectrum at small scales. The fact that HiDM is initially relativistic introduces a characteristic scale:  For the kinetic equilibrium case, this scale corresponds to the sound horizon of HiDM, $l_{\rm{s}}\equiv\int c_{\rm{s}}d\eta$, which determines the scale below which matter overdensity perturbations will be suppressed relative to the standard cold dark matter case.  For the free-streaming case, the suppression scale is instead determined by the free-streaming scale, $l_{\rm{fs}}\equiv\int v d\eta$, where $v$ is the average velocity of HiDM.

Since perturbation evolution closely follows the CDM case after the relativistic-to-nonrelativistic transition, a large-scale mode that enters the horizon after $a=a_{\rm T}^w$ experiences only a small deviation from the standard CDM case. However, this subtle effect still influences the CMB power spectrum, as discussed in Sec.~\ref{section:super-horizon-evolution}. In contrast, a small-scale mode, that enters the horizon before $a=a_{\rm T}^w$, oscillates and experience suppression compared to the standard CDM scenario.

\begin{figure}[tbp]
    \centering
    \includegraphics[width=\textwidth]{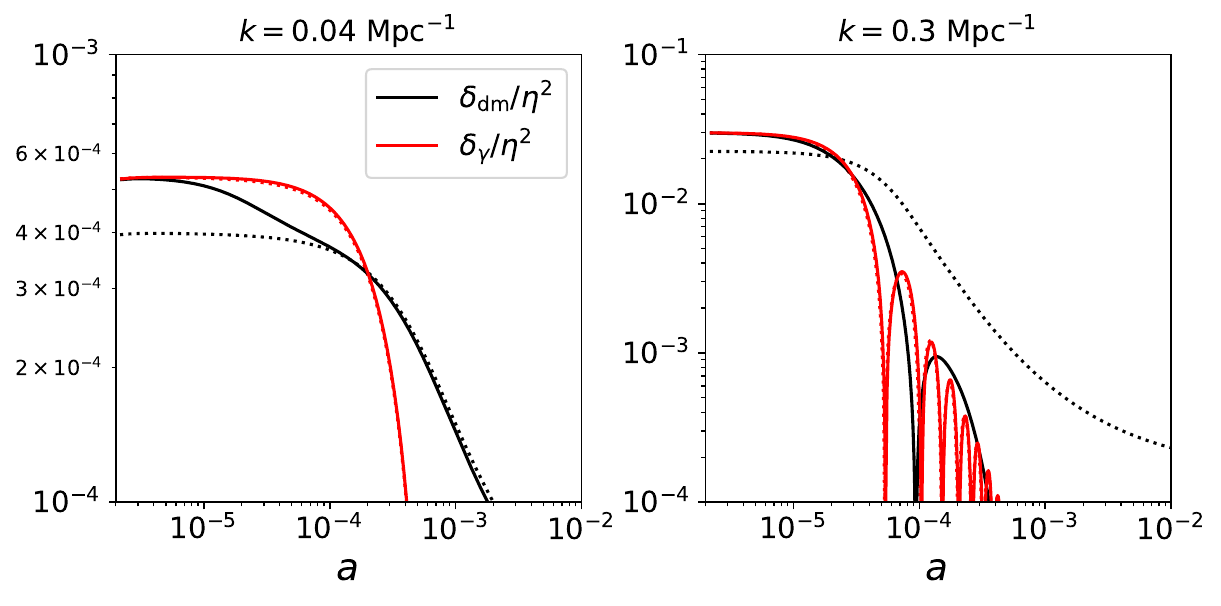}
    \caption{Illustration of the evolution of perturbations over time. Dotted lines are the standard CDM case while solid lines are the HiDM case in kinetic equilibrium. Black and red lines are the DM density and the photon density transfer functions, both of which are divided by the conformal time squared $\eta^2$. Here $R_{\rm{eq}}^{\rm{T}}=0.1$.  For a large-scale mode (small $k$, left panel) that enters the horizon after $a=a_{\rm T}^w$, the dark matter overdensity evolves similarly as the standard CDM case. For a small-scale mode (large $k$, right panel) that enters the horizon before $a=a_{\rm T}^w$, the dark matter overdensity oscillates and get suppressed compared to the standard CDM case. }
    \label{fig:perturbation-evolution}
\end{figure}

In Figure~\ref{fig:perturbation-evolution}, we show two examples for the kinetic equilibrium case: a large-scale perturbation (left panel) and a small-scale perturbation (right panel). By comparing the solid line (HiDM case) and the dotted line (CDM case), we see that the large-scale (small $k$) perturbation in the HiDM case evolves in almost the same way as in the CDM case. The small-scale perturbation (larger $k$) initially follows the photon overdensity before the relativistic-to-nonrelativistic transition. Around and after the transition, the HiDM perturbation deviates from that of photons, but the oscillations persist until the HiDM bulk velocity becomes too small to influence the growth of large-scale structures. At that point, the HiDM overdensity begins to grow via gravitational instability; however, its magnitude today remains significantly smaller than in the standard CDM case.  

In the following sections, we will investigate the HiDM effects at different scales and the constraints from the corresponding observations.

\section{Large scales: impacts on CMB power spectra and constraints from Planck}\label{section:observational-impacts-and-observations-CMB}

\subsection{Super-horizon evolution}\label{section:super-horizon-evolution}
We start from the largest scales and investigate the impacts on the CMB power spectra. We then use the publicly available Planck 2018 baseline data ($TT$, $TE$, and $EE$ power spectra) and perform parameter inference using \textsc{cosmomc} \cite{Cosmomc-Lewis-Bridle-2002}.

\begin{figure}[tbp]
    \centering
    \includegraphics[width=\textwidth]{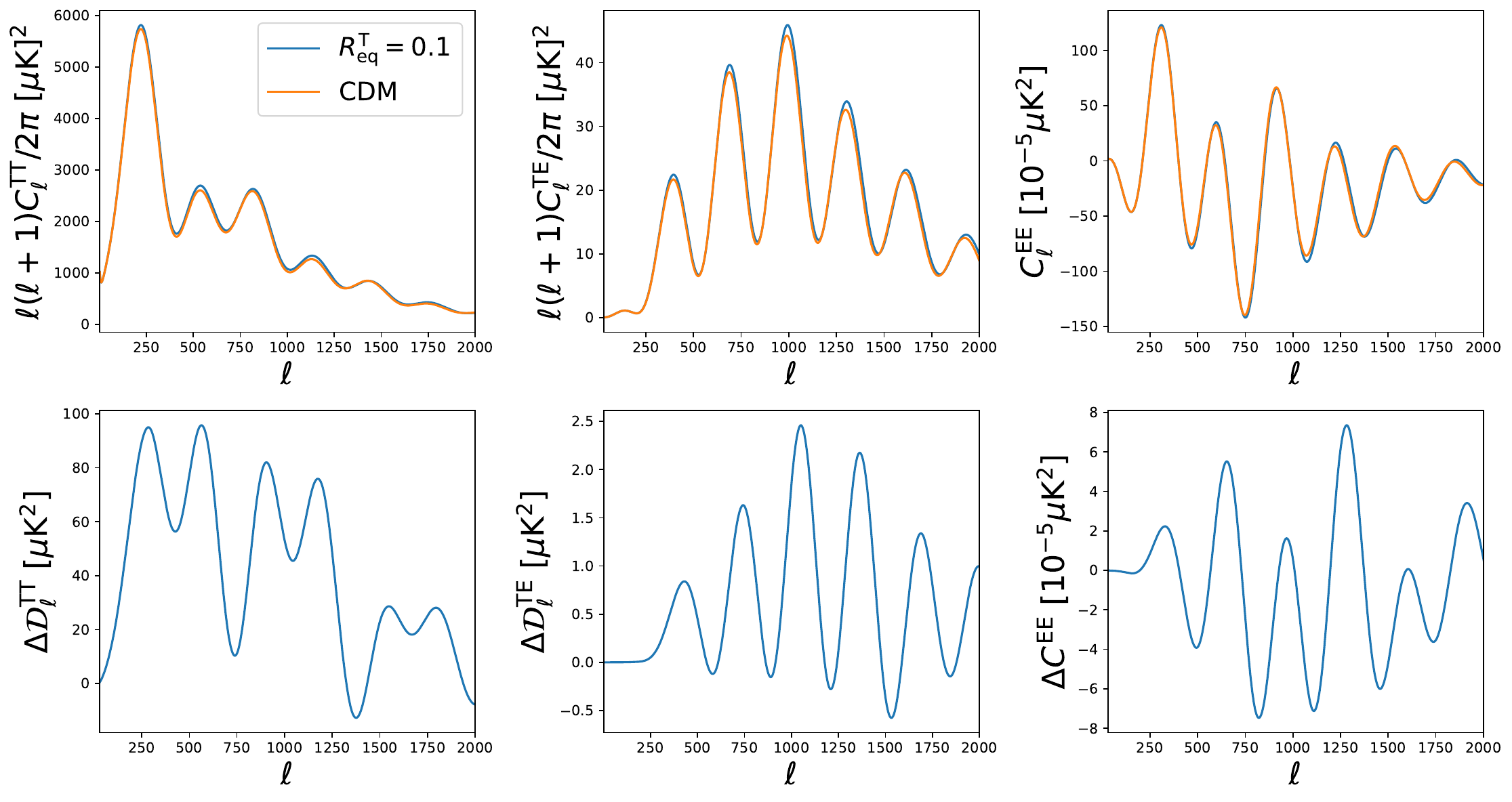}
    \caption{Impacts on the CMB power spectra for the kinetic equilibrium case, where $\mathcal{D}_{\ell}=\ell(\ell+1)C_\ell/2\pi$. To highlight the impact of the HiDM model, we take $R^{\rm{T}}_{\rm{eq}}=0.1$, which is larger than the Planck constraints.}
    \label{fig:CMB-spectra}
\end{figure}

Compared to the small-scale effects that will be detailed later, the impact of HiDM on the CMB power spectra for the same $R^{\rm T}_{\rm eq}$ is relatively small. Consequently, constraints on $R^{\rm T}_{\rm eq}$ from the CMB power spectra alone are expected to be weak. However, for sufficiently large $R^{\rm T}_{\rm eq}$, subtle modulations appear in the temperature anisotropy. Figure~\ref{fig:CMB-spectra} illustrates these effects for a relatively large $R_{\rm eq}^{\rm T}$.

First, as discussed earlier, the matter power spectrum is suppressed at small scales. The suppression scale, $l_{1/2}$, will be quantified in Sec.~\ref{section:observational-impacts-matter-power}. Since the intrinsic CMB temperature fluctuations can be partially canceled by gravitational redshift and blueshift effects, a suppressed matter power spectrum can enhance the CMB temperature power spectrum. However, this effect primarily occurs at scales smaller than those probed by most CMB anisotropies. To quantify this, we take $R^{\rm{T}}_{\rm{eq}}=0.1$ (the value adopted in Figure \ref{fig:CMB-spectra}), which corresponds to a suppression scale of $l_{1/2} \simeq 40$ Mpc [see Eq.~\eqref{eq:ksup-RTeq-numerical}]. The corresponding multipole moment is  
\begin{equation}
\ell \sim \frac{\pi D_{\textsc{a}}^*}{l_{1/2}} \simeq 1100\,,
\end{equation}
where $D_{\textsc{a}}^*\simeq14000$\,Mpc is the comoving angular diameter distance to the last scattering surface. This value of $\ell$ is higher than most of the multipole moments where $\Delta\mathcal{D}_\ell^{\rm TT}$ appears, implying that the enhancement of the CMB temperature power spectrum is driven by a different mechanism than matter power spectrum suppression.

The perturbation modes probed by CMB anisotropies enter the horizon after the relativistic-to-nonrelativistic transition of HiDM. Since their evolution follows the standard case after this transition, the enhancement observed in the CMB temperature power spectrum is primarily governed by differences in their super-horizon evolution compared to $\Lambda$CDM. From Eq.~\eqref{eq:HiDM-initial-adiabatic}, the next-to-leading-order evolution of the photon overdensity, $\delta_\gamma$, is modified, leading to a larger $|\delta_\gamma|$ at a given time.

To estimate this difference, we take the ratio of $\delta_{\gamma}$ between the HiDM and $\Lambda$CDM cases, keeping only the leading-order term for super-horizon modes:
\begin{equation}\label{eq:ratio_delta_gamma}
    \frac{\delta_\gamma^{\rm HiDM}}{\delta_\gamma^{\Lambda{\rm CDM}}}-1\approx\frac{1}{5}\frac{\mathcal{H}_0\eta}{\sqrt{\Omega_{\textsc{r}}}}(\Omega_{\rm m}-\Omega_{\rm b})\,.
\end{equation}
Evaluating this ratio at the transition time, where $\eta\simeq\frac{a_{\rm T}^w}{\mathcal{H}_0\sqrt{\Omega_{\textsc{r}}}}$, gives 
\begin{equation}\label{eq:ratio_numeric}
    \frac{\delta_\gamma^{\rm HiDM}}{\delta_\gamma^{\Lambda{\rm CDM}}}-1\approx\frac{a_{\rm T}^w}{5\Omega_{\textsc{r}}}(\Omega_{\rm m}-\Omega_{\rm b})=\frac{R^{\rm{T}}_{\rm{eq}}}{5}(\Omega_{\rm m}-\Omega_{\rm b})\,.
\end{equation}
For $R^{\rm{T}}_{\rm{eq}}=0.1$, Eq.~\eqref{eq:ratio_numeric} gives a fractional difference of $\sim 0.016$ in $\delta_\gamma$, which translates to a $\sim 0.032$ fractional difference in $\mathcal{D}^{\rm TT}_\ell$. Given that $\mathcal{D}^{\rm TT}_\ell$ ranges from 1000 to 5800 $\mu$K$^2$, the corresponding $\Delta\mathcal{D}^{\rm TT}_\ell$ is between 32 and 180 $\mu$K$^2$. This estimate is consistent with the numerical results shown in Figure~\ref{fig:CMB-spectra} within a factor of 2, indicating differences in the super-horizon evolution of $\delta_\gamma$ are responsible for the effects on CMB power spectra.

Moreover, an increased relativistic energy density accelerates the early expansion of the universe. This faster expansion induces a minor shift in the positions of the power spectrum peaks toward higher multipole values. However, this effect is subdominant.

\subsection{CMB constraints on $R^{\rm{T}}_{\rm{eq}}$}
\begin{figure}[htpb]
    \centering
    \includegraphics[width=\textwidth]{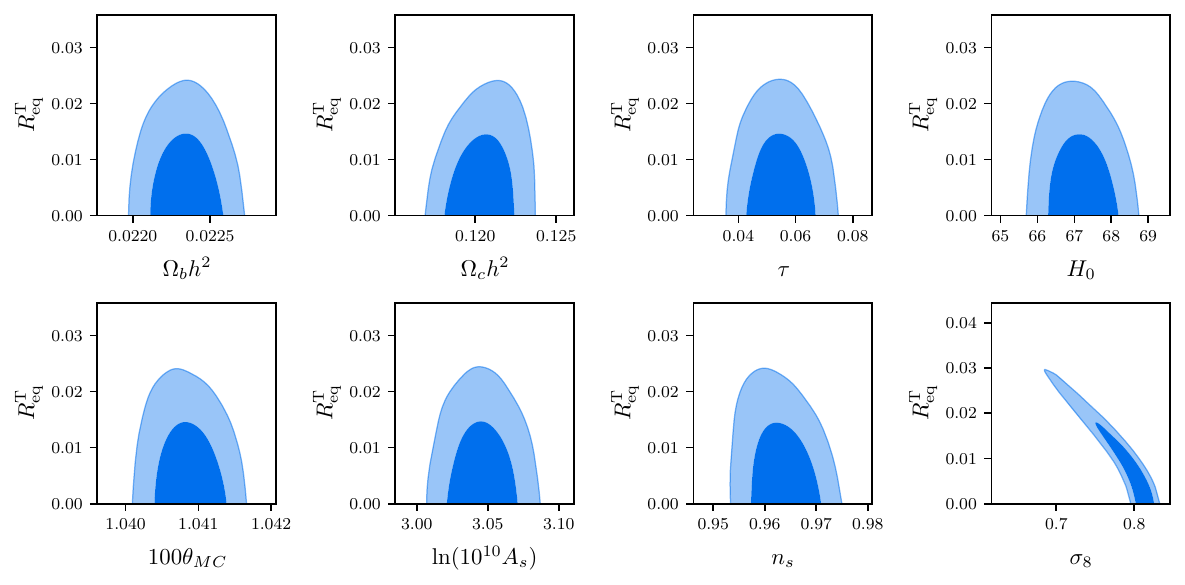}
    \caption{
    Marginalized constraints on $R^{\rm{T}}_{\rm{eq}}$ in the HiDM kinetic equilibrium case versus other standard $\Lambda$CDM model parameters. $R^{\rm{T}}_{\rm{eq}}$ weakly correlates with the six base parameters in the $\Lambda$CDM model. The right column presents two important derived parameters: $H_0$ (which remains unaffected) and $\sigma_8$ (which decreases with increasing $R^{\rm{T}}_{\rm{eq}}$). While this might suggest that HiDM might resolve the $\sigma_8$ tension, constraints from Lyman-$\alpha$ forest data and the smallest halos, as discussed in Sec.~\ref{section:observational-impacts-matter-power}, rule out this possibility.
    \label{fig:constraints_from_CMB}}
\end{figure}

Figure~\ref{fig:constraints_from_CMB} presents the marginalized constraints on $R^{\rm{T}}_{\rm{eq}}$ versus other standard $\Lambda$CDM parameters for both the kinetic equilibrium and free-streaming cases. The value of $R^{\rm{T}}_{\rm{eq}}$ is constrained to $<0.019$ at the $2\sigma$ level. It is evident that $R^{\rm{T}}_{\rm{eq}}$ only weakly correlates with the six base parameters of the $\Lambda$CDM model: $\Omega_{\rm{b}}h^2$, $\Omega_{\rm{c}}h^2$, $\tau$, $100\theta_{\rm{MC}}$, $\ln(10^{10}A_s)$, and $n_s$. Thus, varying $R^{\rm{T}}_{\rm{eq}}$ does not significantly impact the constraints on these standard parameters.

The right panel of Figure~\ref{fig:constraints_from_CMB} also shows constraints on two important derived parameters, $H_0$ and $\sigma_8$. While the constraint on $H_0$ remains largely unaffected, $\sigma_8$ exhibits a correlation with $R^{\rm{T}}_{\rm{eq}}$: larger values of $R^{\rm{T}}_{\rm{eq}}$ lead to lower values of $\sigma_8$. This occurs due to the suppression of small-scale dark matter perturbations. It suggests that adjusting $R^{\rm{T}}_{\rm{eq}}$ could potentially address the $\sigma_8$ tension \cite{Heymans-etal-2012,Hikage-etal-2019,Joudaki-etal-2018}. However, as we show later, Lyman-$\alpha$ forest constraints on $R^{\rm{T}}_{\rm{eq}}$ are an order of magnitude stronger than those from the CMB, eliminating HiDM as a viable solution to the $\sigma_8$ tension.

\section{Intermediate scales, impact on the matter power spectrum, and constraints from the Lyman-Alpha forest}\label{section:observational-impacts-matter-power}

\subsection{The suppression scale of the matter power spectrum}\label{section:suppression-scale}
As discussed in the previous section, the early-time relativistic phase of HiDM suppresses the small-scale matter power spectrum. This suppression effect closely resembles that in the WDM scenario \cite{Bond-Efstathiou-Silk-1980-PhysRevLett,Viel-etal-2005,Bode-2001,2021-Xu-Munoz-Dvorkin}. However, in our HiDM model, we consider both the kinetic equilibrium case (where the suppression scale is set by the sound horizon) and the free-streaming case (where the suppression scale is set by the free-streaming scale like the WDM case).

A more rigorous definition of the suppression scale is based on the ratio of the suppressed dark matter transfer function to the standard CDM transfer function \cite{Viel-etal-2005}. We define the suppression scale as the scale at which the following "transfer function"  
\begin{equation}\label{eq:suppression-scale}
    T(k) \equiv \sqrt{\frac{P_{\Lambda{\rm{HiDM}}}(k)}{P_{\Lambda{\rm{CDM}}}(k)}}
\end{equation}
drops to $1/2$, i.e., $T(k_{1/2})=1/2$. Numerically, we vary $R^{\rm{T}}_{\rm{eq}}$ and use Eq.~\eqref{eq:suppression-scale} to evaluate $k_{1/2}$, finding that $k_{1/2}$ is strongly correlated with $R^{\rm{T}}_{\rm{eq}}$.\footnote{It is worth noting that $k_{1/2}$ does not depend on $H_0$. This is because, given today's radiation energy density, $k_{1/2}$ only depends on $a_{\rm{eq}}$ and $R^{\rm{T}}_{\rm{eq}}$. Thus, $k_{1/2}$ is fully determined when $\Omega_{\rm{m}}h^2$, $\Omega_{\rm{R}}h^2$, and $R^{\rm{T}}_{\rm{eq}}$ are specified.} We adopt the following parameterized relation between $k_{1/2}$ and $a_{\rm T}^w$ (and hence $R^{\rm{T}}_{\rm{eq}}$), as recently proposed in \cite{Zhang:2024vkx}:  
\begin{equation}\label{eq:ksup-RTeq-numerical}
    l_{1/2} \equiv \frac{2\pi}{k_{1/2}} = 1\,{\rm Mpc} \times K\left(\frac{R^{\rm{T}}_{\rm{eq}}}{5.3\times10^{-4}}\right),
\end{equation}
where $\Omega_{\rm{m}}h^2=0.143$ and $\Omega_{\rm{R}}h^2=4.15\times10^{-5}$ have been used, and the function $K(X)$ is defined as  
\begin{equation}
    K(X) = X\left(1 - 0.15 \ln(X)\right).
\end{equation}

The above is for the kinetic equilibrium case. For the free-streaming case with the same $R^{\rm{T}}_{\rm{eq}}$, $l_{1/2}$ is approximately $1.2$ times larger than in the kinetic equilibrium case.

\begin{figure}[tbp]
    \centering
    \includegraphics[width=0.495\textwidth]{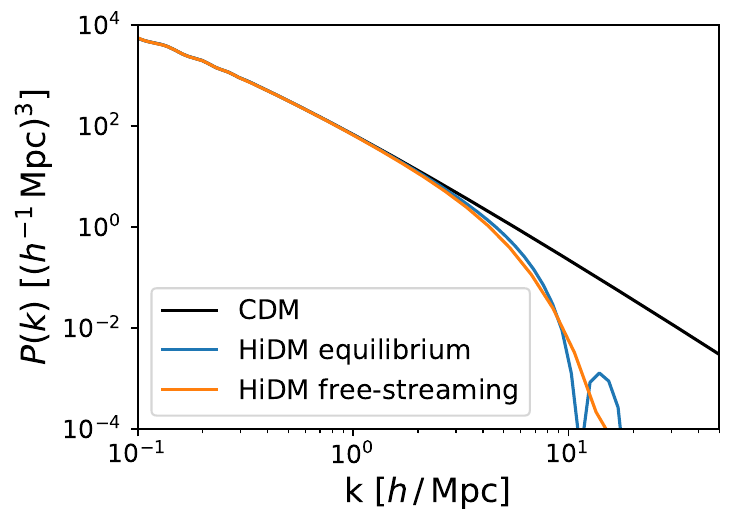}
    \includegraphics[width=0.495\textwidth]{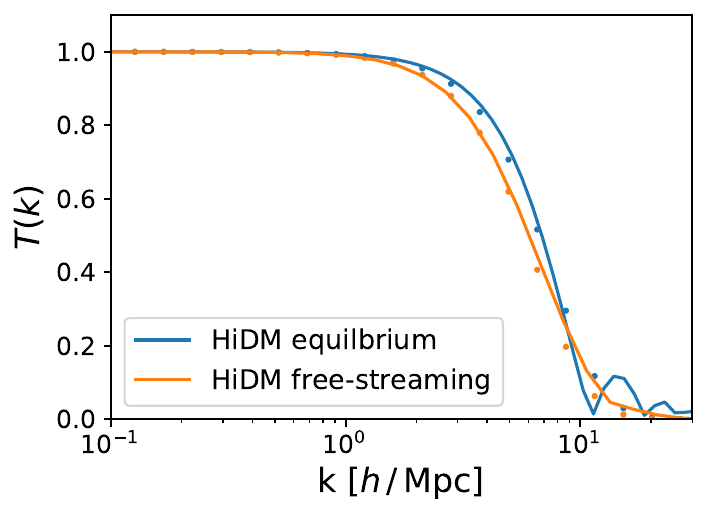}
    \caption{Left: the linear matter power spectrum today in the standard $\Lambda$CDM case (black) and the HiDM case in kinetic equilibrium (blue) and free-streaming (red).  Here $R_{\rm{eq}}^{\rm{T}}=0.001$. The early-time relativistic phase significantly suppresses the power spectrum for small-scale modes with wavenumber larger than some $k_{1/2}$. The suppression in the free-streaming case is more significant. Right: the HiDM ``transfer function'' (note that it is defined in Eq.~\eqref{eq:suppression-scale}, which is different from that shown in Figure~\ref{fig:perturbation-evolution}). The dots are approximations obtained using Eq.~\eqref{eq:Viel-suppression} with the suppression scale $l_{1/2}$ calculated by Eq.~\eqref{eq:ksup-RTeq-numerical}. The suppression scale for the free-streaming case is obtained by multiplying the result of Eq.~\eqref{eq:ksup-RTeq-numerical} by a factor of $1.2$.}
    \label{fig:matter-power}
\end{figure}

As expected, larger values of $R^{\rm{T}}_{\rm{eq}}$ correspond to smaller $k_{1/2}$, shifting the suppression to larger scales. In Figure~\ref{fig:matter-power}, we compare the matter power spectra for CDM and HiDM, including both the kinetic equilibrium and free-streaming cases. The HiDM matter power spectrum is suppressed compared to the CDM case, with the free-streaming case exhibiting a stronger suppression effect than the kinetic equilibrium case for the same value of $R_{\rm eq}^{\rm T}$.

\subsection{Lyman-$\alpha$ forest constraints on $R^{\rm{T}}_{\rm{eq}}$}
As discussed above, the key to constraining $R^{\rm{T}}_{\rm{eq}}$ is to probe the matter power spectrum at small scales. The most effective large-scale cosmological observation that can probe these small scales is the flux power spectrum of the Lyman-$\alpha$ forest \cite{McDonald-etal-2006,Viel-Weller-Haehnelt-2004,Viel-Haehnelt-Lewis-2006}. Since no suppression effects have been detected in the observed matter power spectrum, the suppression wavelength $k_{1/2}$ must be larger than the smallest scales probed by Lyman-$\alpha$ forest observations.  

Recently, Ref.~\cite{Irsic:2023equ} reported that the matter power spectrum cannot decrease by more than 5\% at the wavenumber  
\begin{equation}
    k_{005} > 14.35\,h\,{\rm Mpc}^{-1}.
\end{equation}
To relate this to the suppression scale, we use the empirical fitting function for the WDM transfer function from \cite{Bode-2001}:
\begin{equation}\label{eq:Viel-suppression}
    T(k) \simeq \left(1 + (\alpha k)^{2\mu} \right)^{-5/\mu},
\end{equation}
where $\mu=1.12$ \cite{Viel-etal-2005} and $\alpha$ is given by  
\begin{equation}
    \alpha = \frac{l_{1/2}}{2\pi} \left( 2^{\frac{\mu}{5}} - 1 \right)^{\frac{1}{2\mu}}.
\end{equation}
We also plot this fitting function in the right panel of Figure~\ref{fig:matter-power}, which agrees well with the numerical results. 

Setting $T(k_{005})>14.35\,h$Mpc$^{-1}$ in Eq.~\eqref{eq:Viel-suppression}, we have $l_{1/2} < 0.14$ Mpc, assuming $h=0.7$. Then, from Eq.~\eqref{eq:ksup-RTeq-numerical}, we obtain the following constraints on $R^{\rm{T}}_{\rm{eq}}$:  
\begin{align}
    R^{\rm{T}}_{\rm{eq}} &< 5.5 \times 10^{-5} \quad \text{(kinetic equilibrium case)}, \\
    R^{\rm{T}}_{\rm{eq}} &< 4.5 \times 10^{-5} \quad \text{(free-streaming case)}.
\end{align}
Thus, the Lyman-$\alpha$ forest provides a significantly stronger constraint on $R^{\rm{T}}_{\rm{eq}}$ than CMB.

\section{Small scales: effects on the halo mass function\label{section:HiDM-nonlinear-scales}}
Thus far, our focus has primarily been on the phenomenology and constraints of HiDM at linear scales. In this section, we extend our analysis to nonlinear scales, where the effects become more complex. While a rigorous investigation would require cosmological simulations, we adopt a semi-analytical approach to explore the impact on the halo mass function and the minimum halo mass.

\subsection{The halo mass function}\label{section:HiDM-halo-mass-function}
As discussed in the previous section, the linear power spectrum is suppressed at scales smaller than the sound horizon for the kinetic equilibrium case and the free-streaming scale for the free-streaming case. This suppression affects halo formation at small mass scales and significantly impacts the halo mass function. The characteristic suppressed mass scale in the halo mass function can be estimated as  
\begin{equation}\label{eq:suppressed-mass-scale}
    M_{1/2}\simeq\frac{4\pi}{3}\left(\frac{l_{1/2}}{2}\right)^3\rho_{\rm{m}}^0=4.5\times10^{10}\, M_{\odot}h^2\times \Big[K\big(\frac{R_{\rm{eq}}^{\rm{T}}}{5.3\times10^{-4}}\big)\Big]^3\,.
\end{equation}
The above is for the kinetic equilibrium case. For the free-streaming case, $M_{1/2}$ is larger by a factor of $1.2^3=1.7$ for the same value of $R_{\rm eq}^{\rm T}$. Since the matter power spectrum is suppressed below a specific comoving scale, this estimate of $M_{1/2}$ remains independent of redshift. The halo mass function is expected to be significantly reduced for $M < M_{1/2}$.

To quantify this suppression, we compute the halo mass function using the extended Press-Schechter formalism~\cite{Press-Schechter-1974, Bond-etal-1991-Excursion, Bower-1991, Lacey-etal-1993} with the Sheth-Tormen fitting function~\cite{Sheth-Tormen2002}, given by  
\begin{equation}\label{eq:extended-PS}
    \frac{dn}{dM}(M,z)=A\sqrt{\frac{2a}{\pi}}\frac{\rho_{\rm{m}}(z)}{M}\frac{-d\ln\sigma(M)}{dM}\nu_{\rm{c}}(z)\Big[1+\frac{1}{\big(a\nu^2_{\rm{c}}(z)\big)^q}\Big]\exp\big(-a\nu^2_{\rm{c}}(z)/2\big)\,,
\end{equation}
where $M$ is the halo mass, $\rho_{\rm{m}}(z)$ is the average matter density at redshift $z$, and $\sigma(M)$ is the present-day variance of linear matter fluctuations on scale $M$. The variable $\nu_{\rm{c}}(z)$ is given by 
\begin{equation}\label{eq:nu_c}
    \nu_{\rm{c}}(z) = \frac{1.686}{D(z) \sigma(M)},
\end{equation}
where $D(z)$ is the growth factor normalized to unity today. The fitting parameters are $A = 0.322$, $a = 0.707$, and $q = 0.3$. The variance $\sigma(M)$ is calculated as  
\begin{equation}\label{eq:sigma-R-integral}
\sigma^2(M)=\frac{1}{2\pi^2}\int_0^\infty P_m(k) |f_{\rm sh}(kR)|^2k^2dk\,,
\end{equation}
where the spherical top-hat function $f_{\rm{sh}}(x)$ is,
\begin{equation}\label{eq:spherical-top-hat}
f_{\rm sh}(x)=\frac{3}{x^3}(\sin x-x\cos x)\,,
\end{equation}
and $R=\left(\frac{3M}{4\pi\rho_{\rm{m}}^0}\right)^{1/3}$. While it is common practice to use a sharp-$k$ window function~\cite{Sheth-Tormen2002} to minimize correlated modes in $\sigma^2(M)$, the difference in results is minor. Here, we use the spherical top-hat filter for its clear physical meaning as a spatial averaging within a sphere.

\begin{figure}[tbp]
    \centering
    \includegraphics[width=0.8\textwidth]{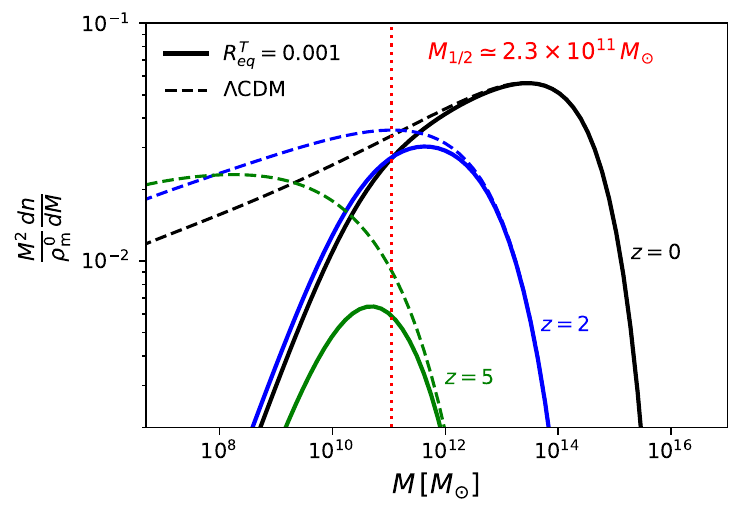}
    \caption{The halo mass functions for the kinetic equilibrium case. A characteristic suppression mass scale, $M_{1/2}$, emerges, below which the number of halos is consistently suppressed compared to the $\Lambda$CDM case. For  $R^{\rm{T}}_{\rm{eq}} = 0.001$, this suppression mass scale is approximately $2.3 \times 10^{11} \, M_{\odot}$. The vertical red-dotted line marks the estimated suppression mass scale $M_{1/2}$, which is calculated from \eqref{eq:suppressed-mass-scale}.}\label{fig:Halo-Mass-function}
\end{figure}

The above formalism allows us to infer the number of collapsed structures at late times from the linear matter power spectrum, which we obtained from our modified \textsc{camb} for the kinetic equilibrium case and \textsc{class} for the free-streaming case. Figure~\ref{fig:Halo-Mass-function} shows the resulting halo mass function for the kinetic equilibrium case, where the suppression scale $M_{1/2}$ estimated from Eq.~\eqref{eq:suppressed-mass-scale} is indicated. For $R^{\rm{T}}_{\rm{eq}} = 0.001$, the number of halos with masses below $M_{1/2} \simeq 2.3 \times 10^{11} M_{\odot}$ (assuming $h = 0.7$) is significantly suppressed, a trend that remains nearly independent of redshift. The results for the free-streaming case are similar, except that $M_{1/2}$ is larger by a factor of $1.7$.

Since larger halos form later in the hierarchical structure formation paradigm, the peak halo mass $M_{\rm{peak}}$ shifts toward lower masses at higher redshifts (see dashed curves in Figure~\ref{fig:Halo-Mass-function}). When $M_{\rm{peak}} < M_{1/2}$, as is the case for $z = 5$, the total number of halos is substantially reduced. This suppression may have important implications for high-redshift phenomena, particularly the epoch of reionization.

It is worth pointing out that the Press-Schechter formalism assumes that the temperature (and thus the pressure or the velocity dispersion) of dark matter is negligible during the collapse. The mass function of the dark matter halos is reduced for $M > M_{1/2}$ due to a suppression of the linear matter power spectrum for $k > k_{1/2}$, which serves as the initial condition for the late-time nonlinear collapse. In our scenario, HiDM has some residual temperature, leading to another effect that suppresses small-halo formation due to changes in the collapse dynamics. The situation is analogous to baryon collapse without dark matter at a finite temperature, where, in the extreme case, a minimum mass scale (the Jeans mass) exists below which the self-gravity of a baryon clump cannot overcome thermal pressure, preventing collapse. Due to the residual temperature of HiDM, there also exists a dark-matter "Jeans mass" ($M_{\rm{J}}$). In Appendix~\ref{section:HiDM-minimum-halo-mass}, we estimate the Jeans mass at different redshifts. We find that it is generally orders of magnitude smaller than $M_{1/2}$, especially for $z < 10$, making it negligible in this context. This agrees with the treatment in Ref.~\cite{2013-Schneider-etal} for warm dark matter.

\subsection{Constraints from the smallest dark matter halo}\label{section:small-halo-constraints}
Recently there have been significant developments in theoretical and observational studies of low-mass galaxies, which are promising targets to test the nature of dark matter \cite{Lin:2016vmm,Nadler-2019-subhalo-DM-B-v-independent-interaction,Ludlow-etal-2016,2013-Benson-etal-DM-halo-beyonCDM,2013-Schneider-etal,Avila_Reese_etal_2001,Maccio-etal-2013,2020-Gilman-etal}. Here we follow Ref. \cite{Nadler-2019-subhalo-DM-B-v-independent-interaction} and use the reported minimum halo mass to perform an order-of-magnitude estimate of the bound on $R_{\rm{eq}}^{\rm{T}}$. We have seen that in the HiDM scenario, there exists a minimum halo mass $M_{1/2}$ today. This $M_{1/2}$ should be at least smaller than any reported halo mass in observation.\footnote{Note that due to baryon effects, such as supernova explosions and other feedbacks, the minimum halo that can host a galaxy may have a mass larger than $M_{1/2}$.} 
By fitting zoom-in simulations to the luminosity distribution of the Milky Way satellites, Ref.\,\cite{Nadler-2019-subhalo-DM-B-v-independent-interaction} reported that  $M_{1/2}<3.1\times10^{8}M_{\odot}$ at $2\sigma$; also see \cite{Jethwa:2016gra}. Similar to \cite{Nadler-2019-subhalo-DM-B-v-independent-interaction}, we set $M_{1/2}<3.1\times10^{8}M_{\odot}$ and use Eq.~\eqref{eq:suppressed-mass-scale} to obtain a constraint of $R_{\rm{eq}}^{\rm{T}}\lesssim7.8\times10^{-5}$ for the kinetic equilibrium case and $R_{\rm{eq}}^{\rm{T}}\lesssim6.4\times10^{-5}$ for the free-streaming case.\footnote{We show in Appendix \ref{section:HiDM-minimum-halo-mass} that the minimum dark halo mass $M_{\rm{J,h}}$ [Eq.~\eqref{eq:HiDM-Jeans-mass}] is generally much lower than $M_{1/2}$. We use $M_{1/2}<5.4\times10^{8}M_{\odot}$ instead of $M_{\rm{J,h}}<5.4\times10^{8}M_{\odot}$ because it is more consistent with the method in used Ref.\,\cite{Nadler-2019-subhalo-DM-B-v-independent-interaction}.}

\section{A List of constraints considered in this work}\label{section:list-of-constraints}
We have examined four different constraints on HiDM, parameterized by $R_{\rm{eq}}^{\rm{T}}$. Each has a different constraining power due to distinct underlying physical mechanisms. Table~\ref{tab:summary-constraints} summarizes the constraints on $R^{\rm{T}}_{\rm{eq}}$ considered in this work. The most stringent constraints currently come from the Lyman-$\alpha$ forest and the small-halo population, both of which provide competitive bounds. 

As discussed in Sec.~\ref{sec:remarks_on_background}, Eq.~\eqref{eq:Om-Req-chi} shows that an upper bound on $R_{\rm{eq}}^{\rm{T}}$ translates into a lower bound on the particle mass, depending on $\Delta_{\rm{ini}}$ and the quantum statistics. For the fiducial case of fermionic HiDM with $\Delta_{\rm{ini}} = 0$, the Lyman-$\alpha$ forest constraint implies a lower bound on the particle mass of $m > 4.7$ keV in the kinetic equilibrium scenario and $m > 5.5$ keV in the free-streaming scenario, assuming $g = 2$. Similarly, the constraint from the small-halo population yields $m > 3.6$ keV for the kinetic equilibrium case and $m > 4.2$ keV for the free-streaming case. These bounds for the free-streaming case are consistent with those reported in \cite{2020-Newton-etal,2020-Enzi-etal,2020-Gilman-etal,Nadler-2019-MW-sattellites}.

Moreover, as Eq.~\eqref{eq:Om-Req-chi} suggests, the mass constraints become more stringent (i.e., higher) for smaller values of $\Delta_{\rm{ini}}$. To relax these bounds, alternative scenarios could be explored, such as a condensation for bosonic HiDM or a positive initial chemical potential for fermionic HiDM~\cite{Zhang:2024vkx}, or the presence of additional species in the dark sector, particularly dark radiation.

\begin{table}[tpb]
    \centering
    {\footnotesize
    \begin{tabular}{clcl}
    \hline\hline
    Observations     & Bounds on $R^{\rm{T}}_{\rm{eq}}$ & Bounds on  $z_{\rm T}^w$ &  \multicolumn{1}{c}{Phenomenological Effects} \\
    \hline
    BBN  & $\lesssim0.065$  & $>5.3\times10^{4}$  & Modifies cosmic expansion during BBN. \\
    Planck 18  & $<0.019$ ($2\sigma$) & $>1.8\times10^{5}$ & Alters CMB power spectra at different scales. \\
    Lyman-$\alpha$   &$<5.5\times10^{-5}$& $>6.2\times10^{7}$  & Suppresses the small-scale matter power spectrum.\\
    Small halos & $\lesssim7.8\times10^{-5}$ & $>4.4\times10^{7}$  & Reduces the population of halos below $M_{1/2}$.\\
    \hline
    \end{tabular}
    \caption{Summary of constraints on $R^{\rm{T}}_{\rm{eq}}$ considered in this work. The constraints on $R_{\rm{eq}}^{\rm{T}}$ are also translated into bounds on the redshift of transition, defined as $z_{\rm T}^w=1/a_{\rm T}^w - 1$. We use ``$\lesssim$'' for estimated constraints. The constraints on $R^{\rm T}_{\rm eq}$ presented here are for the kinetic equilibrium case; for the free-streaming case, the bounds are approximately $0.81$ times smaller.}
    \label{tab:summary-constraints}
    }
\end{table}

\subsection{Remarks on other cosmological/astrophysical phenomenology}
If confirmed, the global 21cm absorption signal~\cite{Bowman-etal-21cm-nature} could provide an additional cosmological constraint on $R_{\rm{eq}}^{\rm{T}}$. The underlying idea is similar to that in Refs.~\cite{Safarzadeh-etal-2018-WDM-21cm,Boyarsky-etal-2019-WDM-21cm,Lopez-etal-2017-WDM-ionization}: suppression of the small-scale matter power spectrum delays the formation of small-mass halos (see also Sec.~\ref{section:HiDM-halo-mass-function}), which in turn postpones star formation and the production of Lyman-$\alpha$ photons. Since the coupling between the spin temperature and gas temperature is mediated by the Lyman-$\alpha$ background, a significant suppression in structure formation could delay this coupling beyond observationally allowed limits, placing constraints on $R_{\rm{eq}}^{\rm{T}}$. However, substantial astrophysical uncertainties remain. These include the early-time star formation rate and the fraction of Lyman-$\alpha$ photons that escape their host halos. Given these uncertainties, deriving a robust 21cm constraint requires detailed astrophysical modeling, which is beyond the scope of this work.

Other astrophysical observations may provide insight into the intrinsic properties of HiDM. In particular, variations in particle mass can influence halo formation and other astrophysical phenomena, such as halo core profiles~\cite{2013-Destri-DeVega-Sanchez-DMHalo,2015-Chavanis-Lemou-Mehats,2015-Chavanis-Lemou-Mehats-II,2015-Chavanis-Lemou-Mehats,Hu-Barkana-Gruzinov-PhysRevLett2000,Burkert_2020,Deng-etal-2018-PhysRevD,Rodrigues-etal-2017} and the signatures of dark matter annihilation or decay~\cite{HESS-DM-anni-2018,AMS-DM-anni-2014,Chan-etal-2019-MD-anni,Depta-etal-2019,Boudaud-etal-2018-DM-constraint,Schon-etal-2018,Schon-etal-2015,Poulin-etal-2016,Enqvist-etal-2015-DM-decay,Blackadder-Koushiappas-2016}. Ref.~\cite{2013-Destri-DeVega-Sanchez-DMHalo} demonstrated that, by considering halo profiles and the minimum halo mass, an upper bound on the mass of single-species fermionic dark matter ($\gtrsim 2$ keV) can be established independently of the cosmological thermal initial conditions (e.g., the initial chemical potential). Additionally, the strength of dark matter self-interactions, which determines whether HiDM is in kinetic equilibrium or free streaming, has been shown to affect halo profiles; see~\cite{2020-Newton-etal,2020-Enzi-etal,Nadler-2019-subhalo-DM-B-v-independent-interaction,Kennedy-Frenk-Carlos-2014,Kaplinghat-Ren-Yu-2020,Tulin-Yu-2018} and references therein. Furthermore, the star formation rate at high redshift has been found to depend on the dark matter mass in the standard warm dark matter (WDM) scenario~\cite{2016-Tan-Wang-Cheng}. This suggests that constraints on HiDM could be further refined through future observations of early star formation.

\section{Summary and conclusion}\label{section:HiDM-summary}
We have explored the cosmological phenomenology of a hidden dark matter model with its own thermal evolution, assuming a conserved comoving particle number and comoving entropy. We examined various scenarios characterized by different quantum statistics, an initial chemical potential that controls particle occupation number, and whether the dark matter remains in kinetic equilibrium or transitions to free streaming. 

Our findings indicate that different cases exhibit similar background evolution as long as the initial chemical potential is nonpositive. This evolution can be well approximated by a set of functions with only one additional parameter beyond the standard $\Lambda$CDM framework: the transition scale factor $a_{\rm T}^w$, which governs the relativistic-to-nonrelativistic transition. For a fixed $a_{\rm T}^w$, the cosmological phenomenology is largely insensitive to the dark matter particle mass, initial temperature, and intrinsic degrees of freedom. It depends only weakly on the initial chemical potential, the particle’s quantum statistics (fermion vs. boson), and whether it remains in kinetic equilibrium or becomes free streaming.

At the perturbation level, we treated the kinetic equilibrium and free-streaming cases separately but found that both yield similar effects on cosmological observables. The CMB power spectra exhibit characteristic modifications, notably an enhancement of temperature anisotropies on large scales. At smaller scales, the matter power spectrum is significantly suppressed—a typical feature when dark matter possesses velocity dispersion or pressure, preventing gravitational collapse. At nonlinear scales, we investigated structure formation by computing the halo mass function using the extended Press-Schechter formalism. Due to the suppression of the matter power spectrum at small scales, the abundance of low-mass halos is significantly reduced across all redshifts. This suppression occurs below a characteristic mass scale $M_{1/2}$, which depends on the transition scale factor.

Finally, we estimated constraints on the transition scale factor from a range of cosmological and astrophysical observations, including BBN, Planck 2018 data, the Lyman-$\alpha$ forest, and the smallest observed subhalo mass. These constraints limit the transition time in different ways, with the most stringent bounds coming from the Lyman-$\alpha$ forest and the observed population of subhalos. Based on the Lyman-$\alpha$ forest constraint, we estimate the transition redshift to be greater than $6.2\times10^7$.  

Together with recent studies in the literature, our results highlight the potential of structure formation at galactic and subgalactic scales as a powerful probe of warm-dark-matter-like models. Looking ahead, upcoming galaxy surveys—including those from LSST\footnote{\href{http://www.lsst.org/lsst}{http://www.lsst.org/lsst}}, DESI\footnote{\href{https://www.desi.lbl.gov/}{https://www.desi.lbl.gov/}}, Euclid\footnote{\href{http://sci.esa.int/euclid/}{http://sci.esa.int/euclid}}, WFIRST\footnote{\href{https://www.skatelescope.org}{https://www.skatelescope.org}}, and JWST\footnote{\href{https://webb.nasa.gov}{https://webb.nasa.gov}}—will provide an unprecedented wealth of data on dwarf galaxies. These datasets will open new avenues to test dark matter models beyond cold dark matter, making this an exciting era for the study of dark matter.

\acknowledgments
W.L. thanks Yuzhu Cui, Fangyuan Gu, Wei Wang, and Zhen Wang for their helpful discussions. 
We thank an anonymous referee for many helpful suggestions that improved this paper.
W. L. acknowledges that this work is supported by the ``Science \& Technology Champion Project" (202005AB160002) and the ``Top Team Project" (202305AT350002), both funded by the ``Yunnan Revitalization Talent Support Program." This work is also supported by the ``Yunnan General Grant'' (202401AT070489). 
K. J. Mack is supported by the National Science Foundation under Grant No. 2108931. We acknowledge the support of the Natural Sciences and Engineering Research Council of Canada (NSERC), RGPIN-03721-2023.
Our calculations and MCMC analyses were performed on the Harvard Odyssey and CfA/iTC clusters as well as the High Performance Computing at NC State University.

\appendix

\section{Details of solving for the background evolution - kinetic equilibrium case}\label{section:background-I-equilibrium}

With the phase space kinetic equilibrium distribution given in Eq.~\eqref{eq:HiDM-distribution}, the particle number density, energy density and pressure are given by,
\begin{align}
n_{\rm{h}}(x,\Delta) &= \frac{gm_{\rm{h}}^3}{2\pi^2\hbar^3}\frac{1}{x^3}\mathcal{N}(x,\Delta)\,,~~\textrm{with~~~}\mathcal{N}(x,\Delta)=\int_0^\infty\frac{dzz^2}{\exp(\sqrt{z^2+x^2}-x-\Delta)\pm1}\,; \label{eq:HiDM-number-density}\\
\rho_{\rm{h}}(x,\Delta) &= \frac{gm_{\rm{h}}^4}{2\pi^2\hbar^3}\frac{1}{x^4}\mathcal{R}(x,\Delta)\,,~~\textrm{with~~~}\mathcal{R}(x,\Delta)=\int_0^\infty\frac{dzz^2\sqrt{z^2+x^2}}{\exp(\sqrt{z^2+x^2}-x-\Delta)\pm1}\,;\label{eq:HiDM-energy-density}\\
p_{\rm{h}}(x,\Delta)&=\frac{gm_{\rm{h}}^4}{2\pi^2\hbar^3}\frac{1}{x^4}\mathcal{P}(x,\Delta)\,,~~\textrm{with~~~}\mathcal{P}(x,\Delta)=\frac{1}{3}\int_0^\infty\frac{dzz^4/\sqrt{z^2+x^2}}{\exp(\sqrt{z^2+x^2}-x-\Delta)\pm1}\,,\label{eq:HiDM-pressure}
\end{align}
with +1 for fermions and -1 for bosons. From Euler's theorem on homogeneous functions, we can obtain the entropy density $s_{\rm{h}}$ as
\begin{equation}\label{eq:HiDM-entropy-density}
    s_{\rm{h}} = \frac{\rho_{\rm{h}}+p_{\rm{h}}-\mu_{\rm{h}} n_{\rm{h}}}{T_{\rm{h}}}\,.
\end{equation}
Then the specific entropy reads
\begin{equation}\label{eq:HiDM-specific-entropy}
    \frac{s_{\rm{h}}}{n_{\rm{h}}}=\frac{\xi(x,\Delta)}{\mathcal{N}(x,\Delta)}-x-\Delta\,,
\end{equation}
where $\xi(x,\Delta)\equiv\mathcal{R}(x,\Delta)+\mathcal{P}(x,\Delta)$.


It is expected that HiDM is relativistic ($x\rightarrow0$) at early times as $a\rightarrow0$ but later becomes nonrelativistic  ($x\rightarrow\infty$). When $x\rightarrow0$, to leading order we have
\begin{align}
    n_{\rm{h}}(x\rightarrow0,\Delta\leq0)&=\frac{gm_{\rm{h}}^3}{2\pi^2\hbar^3}\frac{F_{3}^{\mp}(\Delta)}{x^3}\,,\label{eq:HiDM-number-density-relativistic}\\
    \rho_{\rm{h}}(x\rightarrow0,\Delta\leq0)&=\frac{gm_{\rm{h}}^4}{2\pi^2\hbar^3}\frac{F_{4}^{\mp}(\Delta)}{x^4} \,,\label{eq:HiDM-energy-relativistic}\\
    p_{\rm{h}}(x\rightarrow0,\Delta\leq0)&=\frac{1}{3}\rho_{\rm{h}}(x\rightarrow0,\Delta\leq0)\,,\label{eq:HiDM-pressure-relativistic}\\
\textrm{where~~~~~}    F_s^{\mp}(\Delta)\equiv f^{\mp}_s(\Delta)\Gamma(s)\exp(\Delta)&~~~~~~{\textrm{and,~~~~}}f^{\mp}_s(\Delta)\equiv\sum_{n=0}^\infty\frac{(\mp1)^{n}}{(n+1)^s}\exp(n\Delta)\,,\label{eq:Fs}
\end{align}
where now we have $-$ for fermions and $+$ for bosons. $\Gamma(s)$ above is the gamma function. Some special values of $f_s^{\mp}$ are $f_s^{\mp}(\Delta\ll-1)=1$, $f_s^{+}(0)=\zeta(s)$ and $f_s^{-}(0)=(1-2^{1-s})\zeta(s)$, where $\zeta(s)$ is the zeta function. A useful property of the function  $f_s^{\mp}$ is $\frac{d}{d\Delta}\big( f_s^{\mp}(\Delta)\exp(\Delta)\big)=f_{s-1}^{\mp}(\Delta)\exp(\Delta)$ for $s>2$.  From the above relativistic limits, the solutions satisfying number conservation [$d(na^3)=0$] and the first law of thermodynamics with an adiabatic process [$d(\rho_{\rm{h}}a^3)=-pda^3$] are $x\propto a$ and $\Delta=$ constant. From Eq.~\eqref{eq:HiDM-entropy-density}, we also have
\begin{equation}\label{eq:HiDM-s-relativisitic-limit}
    s_{\rm{h}}(x\rightarrow0,\Delta\leq0)=n_{\rm{h}}\times\Big(\frac{4f^{\mp}_4(\Delta)}{f^{\mp}_3(\Delta)}-\Delta\Big)\,.
\end{equation}

As $a$ increases, HiDM will eventually become nonrelativistic. In the nonrelativistic limit ($x\rightarrow\infty$), to the next leading order we have
\begin{align}
    n_{\rm{h}}(x\rightarrow\infty, \Delta\leq0)&=\frac{gm_{\rm{h}}^3}{2\pi^2\hbar^3}\frac{\sqrt{2}}{x^{3/2}}\Big(F_{3/2}^{\mp}(\Delta)+\frac{5}{4x}F_{5/2}^{\mp}(\Delta)+\mathcal{O}(\frac{1}{x^2})\Big)\,,\label{eq:HiDM-n-non-relativisitc}\\
    \rho_{\rm{h}}(x\rightarrow\infty, \Delta\leq0)&=\frac{gm_{\rm{h}}^4}{2\pi^2\hbar^3}\frac{\sqrt{2}}{x^{3/2}}\Big(F_{3/2}^{\mp}(\Delta)+\frac{9}{4x}F_{5/2}^{\mp}(\Delta)+\mathcal{O}(\frac{1}{x^2})\Big) \label{eq:HiDM-energy-density-non-relativistic} \\
    &= m_{\rm{h}}n_{\rm{h}}\times\left(1+\frac{3}{2x}\frac{f_{5/2}^{\mp}(\Delta)}{f_{3/2}^{\mp}(\Delta)}+\mathcal{O}(\frac{1}{x^2})\right)\,,\notag\\ 
    p_{\rm{h}}(x\rightarrow\infty, \Delta\leq0)&=\frac{gm_{\rm{h}}^4}{2\pi^2\hbar^3}\frac{2\sqrt{2}}{3x^{5/2}}\Big(F_{5/2}^{\mp}(\Delta)+\mathcal{O}(\frac{1}{x})\Big)  \label{eq:HiDM-pressure-non-relativistic}  \\
    &= \frac{m_{\rm{h}}n_{\rm{h}}}{x}\times\left(\frac{f_{5/2}^{\mp}(\Delta)}{f_{3/2}^{\mp}(\Delta)}+\mathcal{O}(\frac{1}{x^2})\right)\,,\notag
\end{align}
again with $-1$ for fermions and $+1$ for bosons. We need to keep terms up to the next leading order here since the leading order of $\rho_{\rm{h}}$ is trivially the rest energy density and it is the next order that reveals the evolution of the x and $\Delta$. From the above nonrelativisitc limits, the solutions satisfying the number conservation and the first law of thermodynamics are $x\propto a^2$ and $\Delta=$ another constant. A much easier way to see why $\Delta$ becomes another constant at the nonrelativistic limit is the following. At the nonrelativistic  limit, plugging Eqs.\,\eqref{eq:HiDM-energy-density-non-relativistic} and \eqref{eq:HiDM-pressure-non-relativistic} into Eq.~\eqref{eq:HiDM-entropy-density} we have 
\begin{equation}\label{eq:s-nonrelativisitic-limit}
    s_{\rm{h}}(x\rightarrow\infty, \Delta\leq0)=n_{\rm{h}}\times\Big(\frac{5}{2}\frac{f_{5/2}^\mp(\Delta)}{f_{3/2}^\mp(\Delta)}-\Delta\Big)\,.
\end{equation}
Since the specific entropy ($s_{\rm{h}}/n_{\rm{h}}$) is a constant, the factor $\big[-\Delta+\frac{5}{2}\frac{f_{5/2}^\mp(\Delta)}{f_{3/2}^\mp(\Delta)}\big]$ must be a constant and so is $\Delta$. By equating the relativistic and nonrelativistic  limits of $s_{\rm{h}}/n_{\rm{h}}$, we can also derive a consistency relation between the initial $\Delta_{\rm{ini}}$ at the relativistic limit and the final $\Delta_{\rm{f}}$ at the nonrelativistic limit,
\begin{equation}\label{eq:Delta-consistency}
    4\frac{f^\mp_4(\Delta_{\rm{ini}})}{f^\mp_3(\Delta_{\rm{ini}})}-\Delta_{\rm{ini}}-\frac{5}{2}\frac{f^\mp_{5/2}(\Delta_{\rm{f}})}{f^\mp_{3/2}(\Delta_{\rm{f}})}+\Delta_{\rm{f}}=0\,.
\end{equation}
In particular, from  $f_s^\mp(\Delta\ll-1)=1$ we have in the classical case that $\lim\limits_{\Delta_{\rm{ini}}\rightarrow-\infty}(\Delta_{\rm{f}}-\Delta_{\rm{ini}})=-1.5$.
We will use the above consistency relation to check our numerical solutions. We shall see that $\Delta$ never increases. So $\Delta\leq0$ at all times if it is initially so. 

From the above analyses, we know that $x$ first increases as $a$ and then as $a^2$, and $\Delta$ gradually changes from one constant to another. Guided by these, we will now derive and solve the differential equations for $x(a)$ and $\Delta(a)$, and then calculate $n_{\rm{h}}(a)$, $\rho_{\rm{h}}(a)$, $p_{\rm{h}}(a)$ and $s_{\rm{h}}(a)$ using Eqs.\,\eqref{eq:HiDM-number-density} to \eqref{eq:HiDM-entropy-density}. The evolution equations for $x$ and $\Delta$ are obtained by keeping the total particle number and the specific entropy of HiDM unchanged, i.e., $d(n_{\rm{h}}a^3)=0$ and $d(s_{\rm{h}}/n_{\rm{h}})=0$.\footnote{Note that we did not use the energy conservation condition $d(\rho_{\rm{h}} a^3)+p_{\rm{h}}d(a^3)=0$. But our consideration is consistent with such a condition, which we have verified with our numerical solutions.} These give the following infinitesimal evolution of $x$ and $\Delta$ in a matrix form,
\begin{align}
    &\begin{pmatrix}d\ln x\\ d\Delta \end{pmatrix} = \bm{M}^{-1} \begin{pmatrix}d\ln a\\ 0 \end{pmatrix}\,,
    \label{eq:differential-x-Delta}\\
    \textrm{with~~~~~~}&\bm{M}\equiv
     \begin{pmatrix}
    1-\frac{x}{3\mathcal{N}}\mathcal{N}_x  &  -\frac{1}{3\mathcal{N}}\mathcal{N}_\Delta \\
    \frac{x}{\mathcal{N}}\xi_x-\frac{x\,\xi}{\mathcal{N}^2}\mathcal{N}_x-x & ~~\frac{1}{\mathcal{N}}\xi_\Delta-\frac{\xi}{\mathcal{N}^2}\mathcal{N}_\Delta-1
    \end{pmatrix} \,, \label{eq:differential-M-matrix}
\end{align}
where the subscript denotes partial derivative (e.g., $\xi_x\equiv\frac{\partial\xi}{\partial x}$) and
\begin{align}
    \mathcal{N}_x&=\int_0^\infty dz\frac{z^2\exp(\sqrt{z^2+x^2}-x-\Delta)}{\big[\exp(\sqrt{z^2+x^2}-x-\Delta)\pm1\big]^2}\Big(1-\frac{x}{\sqrt{z^2+x^2}}\Big)\,, \label{eq:Nx}\\
    \mathcal{N}_\Delta&=\int_0^\infty dz\frac{z^2\exp(\sqrt{z^2+x^2}-x-\Delta)}{\big[\exp(\sqrt{z^2+x^2}-x-\Delta)\pm1\big]^2}\,,\label{eq:NDelta}\\
    \xi_x&=\int_0^\infty dz \frac{z^2}{\exp(\sqrt{z^2+x^2}-x-\Delta)\pm1}\Bigg[\frac{(\sqrt{z^2+x^2}-x)(1+\frac{z^2}{3(z^2+x^2)})}{1\pm\exp(-\sqrt{z^2+x^2}+x+\Delta)}\nonumber\\
    &~~~~~~~~~~~~~~~~~~~~~~~~~~~~~~~~~~~~~~~~~~~~~~+\frac{x}{\sqrt{z^2+x^2}}-\frac{xz^2}{3(z^2+x^2)^{3/2}}\Bigg]\,,\label{eq:xix}\\
    \xi_{\Delta}&=\int_0^\infty dz\frac{z^2\exp(\sqrt{z^2+x^2}-x-\Delta)}{\big[\exp(\sqrt{z^2+x^2}-x-\Delta)\pm1\big]^2}\Big[\sqrt{z^2+x^2}+\frac{z^2}{3\sqrt{z^2+x^2}}\Big]\,.\label{eq:xiDelta}
\end{align}
Given the initial conditions of $x$ and $\Delta$, the above equations are solved by integrating from a very small $a$ when $x\ll1$ to $a=1$. We can also write Eq.~\eqref{eq:differential-x-Delta} in the form of decoupled differential equations
\begin{equation}\label{eq:lnx-Delta-ODE}
    \frac{d\bm{y}}{d\ln a}=\bm{f}\,,
\end{equation}
where $\bm{y}=(\ln x,\,\Delta)$ and $f_i=(\bm{M}^{-1})_{i1}$.

By solving the evolution of $x$ and $\Delta$, we can also obtain the (squared) adiabatic sound speed squared $c_{\rm{s}}^2$ that is used at the perturbation level. This can be done by restricting $\delta x$ and $\delta \Delta$ to the path (in the state variable space) that keeps the specific entropy unchanged. For matter that has only two independent state variables  (e.g., $T_{\rm{h}}$ and $\mu_{\rm{h}}$ here), this perturbation path is unique and is $(\delta x,\delta \Delta) \propto \big(M_{22},-M_{12}\big)$. But since we also assume the background evolution is keeping the specific entropy unchanged, all state variables' infinitesimal time variations must be along the same path. All perturbed quantities ($\delta x$, $\delta \Delta$, $\delta \rho_{\rm{h}}$, $\delta p_{\rm{h}}$) must then be proportional to their infinitesimal temporal variations ($dx$, $d\Delta$, $d\rho_{\rm{h}}$, $dp_{\rm{h}}$). Therefore, there are two methods of obtaining the adiabatic sound speed, which are
\begin{equation}
\begin{split}
    \frac{\delta p_{\rm{h}}}{\delta \rho_{\rm{h}}}&=\frac{\frac{\partial p_{\rm{h}}}{\partial x}\delta x+\frac{\partial p_{\rm{h}}}{\partial \Delta}\delta\Delta}{\frac{\partial \rho_{\rm{h}}}{\partial x}\delta x+\frac{\partial \rho_{\rm{h}}}{\partial \Delta}\delta\Delta} 
    \\ &= \frac{(4-x\mathcal{P}_x)M_{22}+\mathcal{P}_\Delta M_{12}}{(4-x\mathcal{R}_x)M_{22}+\mathcal{R}_\Delta M_{12}}\,, \textrm{~~~(method 1)}\\
    &=\frac{dp_{\rm{h}}}{d\rho_{\rm{h}}} \rm{~~or~~} \frac{d p_{\rm{h}}/da}{d\rho_{\rm{h}}/da}\,.\textrm{~~~~(method 2)}\label{eq:adiabatic-sound-speed-usual-form}
\end{split}
\end{equation}
Method 1 is valid if the perturbations are isentropic, while method 2 is valid if the perturbations and the background evolutions are both isentropic. We have verified the equivalence of the above two methods when both the background evolution and the perturbations are isentropic. The evolution of the adiabatic sound speed is also shown in the lower middle panel of Figure \ref{fig:sample_result}. We can see that $\delta p_{\rm{h}}/\delta \rho_{\rm{h}}$ has a very similar scale-factor dependence as $w_{\rm{h}}$ but with a delayed transition time.

\section{Details of finding a parameterized background evolution}\label{section:exact-parameterized}

\subsection{A parameterized background evolution}\label{section:background-parameterized}

\begin{figure}[tbp]
    \centering
    \includegraphics[width=0.8\textwidth]{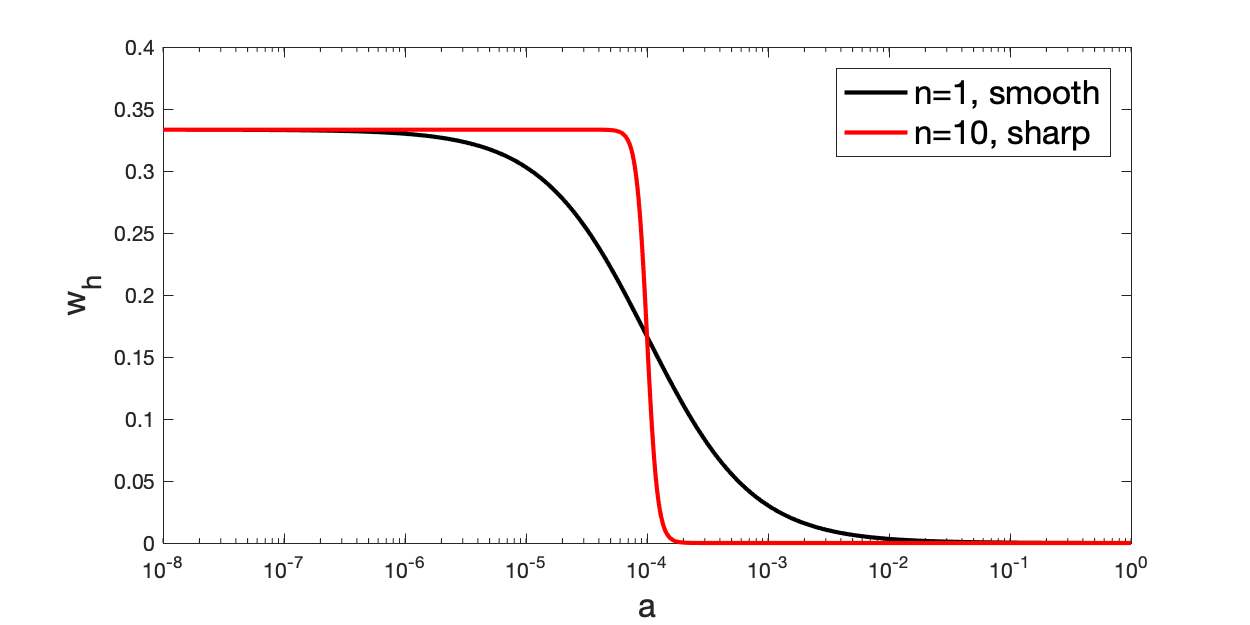}
    \caption{Examples of a relatively smooth (n=1) and a relatively sharp (n=10) transition described by a parameterized equation-of-state parameter $w_{\rm{h}}(a)$ given by in Eq.~\eqref{eq:HiDM-w-parameterization}. }
    \label{fig:transition-parameterized}
\end{figure}

The hidden dark matter began relativistic, with a temperature $T_{\rm{h}}$ different from that of the Standard Model particles. After $T_{\rm{h}}$ dropped below the mass of the hidden baryon ($m_{\rm{h}}$), it transitioned to the nonrelativistic case. We find that the evolution of the equation-of-state parameter $w_{\rm{h}}(a)$ can be well approximated by the following parameterized function,
\begin{equation}\label{eq:HiDM-w-parameterization}
    w_{\rm{h}}(a)=\frac{1}{3}\frac{a_{\rm{T}}^n}{a^n+a_{\rm{T}}^n}\,,
\end{equation}
where $a_{\rm{T}}$ specifies the transition scale factor and the positive transition index $n$ determines the transition width. The equation-of-state parameter approaches $1/3$ for $a\ll a_{\rm{T}}$ and $0$ for $a\gg a_{\rm{T}}$. The transition is shallower (sharper) for a smaller (larger) $n$. Note that we do not restrict $n$ to be an integer. With the above parameterized $w_{\rm{h}}$, the energy density can be calculated analytically and reads,
\begin{equation}\label{eq:HiDM-density-parameterization}
    \rho_{\rm{h}}(a)=\frac{\rho_{\rm{h}}^0}{a^4}\left(\frac{a^n+a_{\rm{T}}^n}{1+a_{\rm{T}}^n}\right)^{\frac{1}{n}}\,,
\end{equation}
where the superscript or subscript ``0'' denotes today and we normalize the scale factor today to unity $a_0=1$. We require that dark matter is highly nonrelativistic today, $a_{\rm{T}}\ll1$. For $n$ of the order of unity, we have $\rho_{\rm{h}}\simeq\rho_{\rm{h}}^0/a^3$ at late times when $a\gg a_{\rm{T}}$ and $\rho_{\rm{h}}\simeq\rho_{\rm{h}}^0a_{\rm{T}}/a^4$ at early times when $a\ll a_{\rm{T}}$. Examples of the parameterized equation of state parameter are shown in Figure \ref{fig:transition-parameterized}.

At the perturbation level, we will also need the squared sound speed $c_{\rm{s}}^2\equiv\delta p_{\rm{h}}/\delta \rho_{\rm{h}}$. If we assume that the specific entropy is unchanged for both the background evolution and perturbations of HiDM, the sound speed can be calculated as
\begin{equation}\label{eq:sound-speed-parameterized}
\begin{split}
    \frac{\delta p_{\rm{h}}}{\delta \rho_{\rm{h}}}&=\frac{dp_{\rm{h}}/da}{d\rho_{\rm{h}}/da} \,, \textrm{~~~~~~~~~if both background and perturbations are isentropic,}  \\
    &= \frac{1}{3}\frac{\tfrac{4}{3}a_{\rm{T}}^n}{a^n+\tfrac{4}{3}a_{\rm{T}}^n}\,,\textrm{~~~~~using the parameterized } w_{\rm{h}}(a)\textrm{ from Eq.~\eqref{eq:HiDM-w-parameterization}.}
\end{split}
\end{equation}
Note that $\delta p_{\rm{h}}/\delta \rho_{\rm{h}}$ has the same functional dependence on $a$ as $w_{\rm{h}}(a)$ with the same transition index but a larger transition scale factor $a_{\rm{T}}'=(\frac{4}{3})^{\tfrac{1}{n}}\times a_{\rm{T}}$. On a logarithmic scale, the transition ``gap'' is $\log_{10}a_{\rm{T}}'-\log_{10}a_{\rm{T}}=\frac{1}{n_w}\log_{10}(4/3)$.

We will fit Eq.~\eqref{eq:HiDM-density-parameterization} into the numerical result of $\rho_{h}$ to determine the best-fit transition index $n_w$ and transition scale $a_{\rm{T}}^w$. These $n_w$ and $a_{\rm{T}}^w$ will be used in Eq.~\eqref{eq:HiDM-w-parameterization} to approximate the evolution of $w_{\rm{h}}$. For $c_{\rm{s}}^2$, since Eq.~\eqref{eq:sound-speed-parameterized} has the same form as Eq.~\eqref{eq:HiDM-w-parameterization}, instead of using Eq.~\eqref{eq:sound-speed-parameterized} we use another Eq.~\eqref{eq:HiDM-w-parameterization} with a different set of $n_{c_{\rm{s}}^2}$ and $a_{\rm{T}}^{c_{\rm{s}}^2}$ to fit $c_{\rm{s}}^2(a)$ for a better approximation to the numerical result. According to the above discussions, we should have $n_{c_{\rm{s}}^2}\simeq n_w$ and $a_{\rm{T}}^{c_{\rm{s}}^2}\simeq (\frac{4}{3})^{1/n_w}\times a_{\rm{T}}^w$; we will use these conditions to check our results.

\subsection{Parameterized approximation I: the kinetic equilibrium case}\label{section:approximation-I-equilibrium}
Let us see what can affect the background evolution. In the classical limit ($\Delta\ll-1$), the transition profile is independent of $\Delta_{\rm{ini}}$. All other parameters associated with the particle nature, e.g. $g$ and $m$, can be absorbed into a single parameter: the HiDM energy density fraction today $\Omega_{\rm{h}}$. Therefore, when $\Delta_{\rm{ini}}\ll-1$ the evolution of energy density and equation of state of HiDM is determined by only two parameters: $\Omega_{\rm{h}}$ and $R_{\rm eq}^{\rm T}$. When $\Delta_{\rm{ini}}\gtrsim-1$, quantum statistics plays a role in the evolution. Let us take the equation of state parameter $w_{\rm{h}}$ as an example. It does not depend on $\Omega_{\rm{h}}$ and generally we can write it as a function of $a$ that depends on the initial warmness, $\Delta_{\rm{ini}}$, and the underlying statistics,
\begin{equation}\label{eq:w_h-evolution-general}
    w_{\rm{h}}=w_{\rm{h}}(a;T_{\rm h}^0/m_{\rm h},\Delta_{\rm{ini}},\pm)\,,
\end{equation}
where ``$+$'' here stands for Fermi-Dirac statistics and ``$-$'' for Bose-Einstein statistics. We define the classical limit of $w_{\rm{h}}$ as
\begin{equation}\label{eq:w_h-classical-limit}
    w_{\rm{h}}^{\rm{class.}}(a;T_{\rm h}^0/m_{\rm h})\equiv w_{\rm{h}}(a;T_{\rm h}^0/m_{\rm h},\Delta_{\rm{ini}}\ll-1,\pm)\,.
\end{equation}
We shall show in Sec.\,\ref{section:dependence-on-Delta} that its dependence on $\Delta_{\rm{ini}}$ or the underlying statistics is rather weak. Similarly, the evolution of other variables can be written as $x=x(a;T_{\rm h}^0/m_{\rm h},\Delta_{\rm{ini}},\pm)$, $\rho_{\rm{h}}=\rho_{\rm{h}}(a;\Omega_{\rm{h}},T_{\rm h}^0/m_{\rm h},\Delta_{\rm{ini}},\pm)$, $p_{\rm{h}}=p_{\rm{h}}(a;\Omega_{\rm{h}},T_{\rm h}^0/m_{\rm h},\Delta_{\rm{ini}},\pm)$ and $\delta p_{\rm{h}}/\delta\rho_{\rm{h}}=c_{\rm{s}}^2(a;T_{\rm h}^0/m_{\rm h},\Delta_{\rm{ini}},\pm)$, etc. They all have classical limits that only depend on $T_{\rm h}^0/m_{\rm h}$, or in addition, on $\Omega_{\rm{h}}$; see Sec.\,\ref{section:dependence-on-Delta}.

The background evolution can be very well approximated by the parameterization presented in Sec.\,\ref{section:background-parameterized}. To do this, we look for the $\rho_{\rm{h}}^{\rm{approx}}$ [parameterized by $n$ and $a_{\rm{T}}$ in Eq.~\eqref{eq:HiDM-density-parameterization}] that best matches the numerical solution of $\rho_{\rm{h}}$. Note that we choose to match $\rho_{\rm{h}}$ instead of $w_{\rm{h}}$ because the energy density is more physically important than the equation of state parameter. Our best-fit $\rho_{\rm{h}}^{\rm{approx}}$ is defined to minimize the following difference:
\begin{equation}\label{eq:difference-rho_h-approximation}
    {\rm{Difference}}=\int d\ln(a) \big[\ln \rho_{\rm{h,N}}(a;T_{\rm h}^0/m_{\rm h},\Delta_{\rm{ini}},\pm) - \ln\rho_{\rm{h,N}}^{\rm{approx}}(a;n_w,a_{\rm T}^w)\big]^2\,,
\end{equation}
where the normalized energy densities are defined as $\rho_{\rm{h,N}}\equiv\frac{\rho_{\rm{h}}}{gm^4/(2\pi^2\hbar^3)}$ and $\rho_{\rm{h,N}}^{\rm{approx}}\equiv\frac{\rho_{\rm{h}}^{\rm{approx}}}{gm^4/(2\pi^2\hbar^3)}$ (so that the $\Omega_{\rm{h}}$ dependence is canceled out). The superscript/subscript $w$ denotes the fitting parameters for $\rho_{\rm{h}}$ and $w_{\rm{h}}$, which are to be distinguished from the parameters for the sound speed later. Using Eq.~\eqref{eq:HiDM-density-parameterization} and imposing $\rho_{\rm{h}}^{\rm{approx}}(a\rightarrow0)=\rho_{\rm{h}}(a\rightarrow0)$, we obtain the normalized density fitting function as
\begin{equation}\label{eq:HiDM-rho-approximation-normalized}
    \rho_{\rm{h,N}}^{\rm{approx}}(a)=\big(\lim\limits_{a\rightarrow0}\frac{a}{x}\big)^4 \frac{F_4^\mp(\Delta_{\rm{ini}})}{a^4}\big[1+(a/a_{\rm{T}}^w)^{n_w}\big]^{\frac{1}{n_w}}\,.
\end{equation}
We vary $n_w$ and $a_{\rm{T}}^w$ to minimize Eq.~\eqref{eq:difference-rho_h-approximation}. Once the best-fit $n_{w}$ and $a_{\rm{T}}^w$ are found, we calculate $w_{\rm{h}}^{\rm{approx}}$ with Eq.~\eqref{eq:HiDM-w-parameterization} to approximate the numerical evolution of $w_{\rm{h}}$. We show in the left panel of Figure \ref{fig:approx-w-dpdrho} the fractional difference between some examples of the numerical evolution of the energy density and its best-fit approximation. In this example, we take a fermionic HiDM with $\Delta_{\rm{ini}}=0$ and $\lim\limits_{x\rightarrow0}\frac{x}{a}=10^4$. The difference between the numerical solution and the approximation is within $0.5\%$ at all times. In the middle panel, we show the numerical solution of 
$w_{\rm{h}}$ (solid black) of the same example and its approximation $w_{\rm{h}}^{\rm{approx}}$ [dashed red, calculated with Eq.~\eqref{eq:HiDM-w-parameterization}]. They also match very well. Table \ref{tab:transition-parameters} lists the best-fit parameters for some special cases. We find that the transition index is about $n_w\simeq1.8$ for all cases of interest, which means the transition is rather smooth. 

The next step is to determine the evolution of the adiabatic sound speed $c_{\rm{s}}^2$. We could use Eq.~\eqref{eq:sound-speed-parameterized} to calculate it with the same $n_w$ and $a_{\rm{T}}^w$. However, we pointed out earlier that Eq.~\eqref{eq:sound-speed-parameterized} has the same functional dependence on $a$ as Eq.~\eqref{eq:HiDM-w-parameterization}. For a closer approximation to the numerical solution, it is better to fit the evolution of the sound speed using another fitting function $c_{\rm{s,approx}}^2$ of the form of Eq.~\eqref{eq:HiDM-w-parameterization} but with another set of $n_{c_{\rm{s}}^2}$ and $a_{\rm T}^{c_{\rm{s}}^2}$. That is, we look for the approximation $c_{\rm{s,approx}}^2$ that minimizes the following difference,
\begin{align}
    {\rm{Difference}'}&=\int d\ln(a) \big[c_{\rm{s}}^2(a;T_{\rm h}^0/m_{\rm h},\Delta_{\rm{ini}},\pm) - c_{\rm{s,approx}}^2(a;n_{c_{\rm{s}}^2},a_{\rm T}^{c_{\rm{s}}^2})\big]^2\,,\label{eq:difference-cs2-approximation}\\
    {\rm{with}}&~~~~~c_{\rm{s,approx}}^2=\frac{1}{3}\frac{1}{(a/a_{\rm{T}}^{c_{\rm{s}}^2})^{n_{c_{\rm{s}}^2}}+1}\,.
\end{align}
The superscript/subscript $c_{\rm{s}}^2$ denotes the fitting parameters for the sound speed, which are different from those for $\rho_{\rm{h}}$ and $w_{\rm{h}}$. The best-fit parameters for the sound speed are also listed in Table \ref{tab:transition-parameters}. We find that the best-fit transition index is about $n_{c_{\rm{s}}^2}=1.9$ for all cases. So, the sound speed transition is slightly sharper than the equation of state transition. 


For the kinetic equilibrium case, we can also determine the values of $\hat{a}$ and $C$ according to their definitions in Eqs.\,\eqref{eq:x-relativistic-limit} and \eqref{eq:x-nonrelativisitic-limit}. The constant $C$ can be determined from the conservation of comoving number density and the fact that $C=\big(\lim\limits_{x\rightarrow0}\frac{x}{a}\big)^2\big/\big(\lim\limits_{x\rightarrow\infty}\frac{x}{a^2}\big)$. From  Eqs.\,\eqref{eq:HiDM-number-density-relativistic} and \eqref{eq:HiDM-n-non-relativisitc} we have
\begin{equation}\label{eq:C-consistency-appendix}
    C=\frac{2}{\pi^{1/3}}\Big[\frac{f_3^\mp(\Delta_{\rm{ini}})}{f_{3/2}^\mp(\Delta_{\rm{f}})}\Big]^{\frac{2}{3}}\exp\big(\frac{2}{3}(\Delta_{\rm{ini}}-\Delta_{\rm{f}})\big)\,,
\end{equation}
where $\Delta_{\rm{f}}$ can be calculated from Eq.~\eqref{eq:Delta-consistency} for a given $\Delta_{\rm{ini}}$. In particular, the classical limit of $C$ is $\lim\limits_{\Delta_{\rm{ini}}\rightarrow-\infty}C=\big(\frac{8}{\pi}\big)^{\frac{1}{3}}\exp(1)\simeq3.7120$. Knowing the value of $C$ is convenient because we can easily compute the temperature-to-mass ratio in the nonrelativistic limit when given its relativistic limit. Some special values of $C$ are listed in Table \ref{tab:transition-parameters}. 

\subsection{Parameterized approximation II: the free-stream case} \label{section:approximation-fs}
We have shown in Sec.\,\ref{section:background-fs-case} that the background evolution of the free-streaming case is very similar to the kinetic equilibrium case. Thus, it is expected that the parameterized functions [Eqs.\,\eqref{eq:HiDM-w-parameterization} and \eqref{eq:HiDM-density-parameterization}] can well approximate the background evolution of the free-streaming case, and we shall see that it does. 

For the free-streaming case, there is no $\hat{a}$ that can be defined using Eqs.\,\eqref{eq:x-relativistic-limit} and Eq.~\eqref{eq:x-nonrelativisitic-limit}, since the distribution is no longer in equilibrium. But the parameter $a_{\rm{T}}$ in Eq.~\eqref{eq:HiDM-w-parameterization} still determines the time (in $a$) of the transition. Similar to the kinetic equilibrium case, we use Eq.~\eqref{eq:HiDM-density-parameterization} to fit the numerical solution of $\rho_{\rm{h}}$ to obtain the best-fit parameters $n_{w}$ and $a_{\rm{T}}^w$ and calculate $w_{\rm{h}}$ with Eqs\,\eqref{eq:HiDM-w-parameterization} using the same $n_{w}$ and $a_{\rm{T}}^w$. 

\begin{figure}[tbp]
    \centering
    \includegraphics[width=0.6\textwidth]{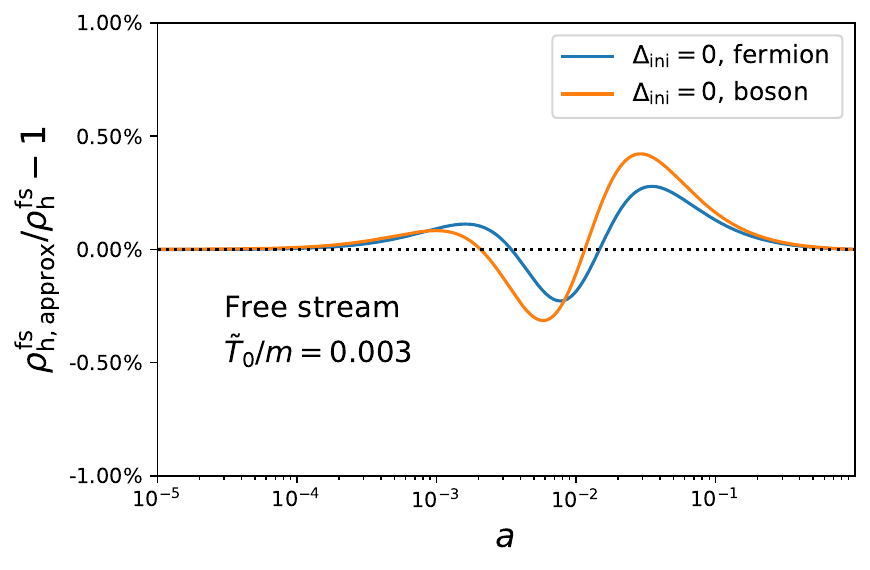}
    \caption{Comparision of the approximation and the numerical solution of the energy density evolution in the free-stream case. The difference is within about $0.5\%$ for both fermion and boson cases at all times.}
    \label{fig:approximation-tonumerical-fs}
\end{figure}
We show in Figure \ref{fig:approximation-tonumerical-fs} the fractional difference between the approximation and the numerical solution of the energy density for both the fermion and boson cases. The approximation is also very good for the free-streaming case, and the fractional difference is within about $0.5\%$ for both cases at all times. Interestingly, compared to the corresponding cases in kinetic equilibrium, we can see that the free-streaming fermion and boson cases have slightly smaller transition indices. This means that it takes slightly longer for free streaming models to finish the relativistic-to-nonrelativistic transition.

It is worth to note that the suppression of the small-scale matter power spectrum observed here differs from that reported in Ref.~\cite{2018Das-etal-Ballistic-DM}. As shown in their Figure 3, the matter power spectrum amplitude in their model can either be suppressed or enhanced. To investigate this discrepancy, we artificially varied the transition index of the equation of state. We found that an enhancement of the matter power spectrum at small scales occurs only if the relativistic-to-nonrelativistic transition is sufficiently sharp (i.e., if $n_w>5$). However, in our model—where the evolution is derived from first principles based on realistic dark matter particle properties—the transition is relatively smooth, with $n_w\sim2$.

\section{The weak dependence on the initial chemical potential or the underlying statistics}\label{section:dependence-on-Delta}

\begin{figure}[tbp]
    \centering
    \includegraphics[width=\textwidth]{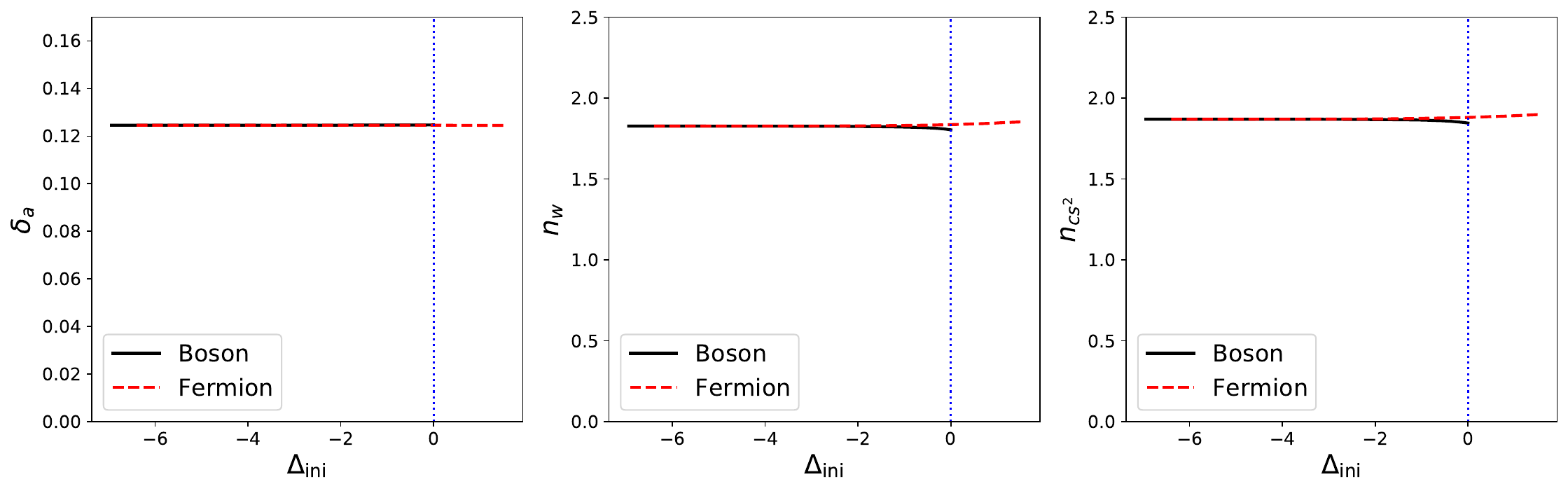}
    \caption{The dependence of the transition parameters on $\Delta_{\rm{ini}}$ and Fermi-Dirac/Bose-Einstein statistics. Although we only consider cases where $\Delta\leq0$, we include some positive $\Delta_{\rm{ini}}$ for the fermion case. Left: the ``gap'' in the transition scale factors in the logarithmic scale between the equation of state and the sound speed. Middle and right: the transition indices of the equation of state and the sound speed. These transition parameters are insensitive to $\Delta_{\rm{ini}}$ or different quantum statistics, justifying the use of the transition scale factor as the only parameter to characterize the HiDM model.
    }
    \label{fig:F-B_transitions}
\end{figure}
 
We can see from Table \ref{tab:transition-parameters} that the transition depends on the initial effective chemical potential $\Delta_{\rm{ini}}$ and whether HiDM is fermion or boson. We will now investigate such dependence in more detail. We shall however find that the classical approximation is already very good at representing the transition in all cases.

In Figure \ref{fig:F-B_transitions}, we show the transition parameters as functions of $\Delta_{\rm{ini}}$ for the kinetic equilibrium case. The transition “gap” between the equation of state and the sound speed remains nearly constant. The transition indices, $n_w$ and $n_{cs^2}$, vary only slightly for $\Delta_{\rm{ini}} \gtrsim -1$.
For the fermion case, the additional degenerate pressure causes the transition to become slightly sharper, reflected in a larger transition index. Conversely, for the boson case, the lower pressure compared to the classical case results in the opposite effect, though the changes remain minimal. Both transition indices are largely insensitive to $\Delta_{\rm{ini}}$ and quantum statistics.
For the cosmological observations discussed in this work, the key time-dependent quantities are $w_{\rm{h}}$ and $\delta p_{\rm{h}} / \delta \rho_{\rm{h}}$. 

We have shown the similarity at the background level between the kinetic equilibrium case and the free-streaming case in Sec.\,\ref{section:background-fs-case} and \ref{section:approximation-fs}. Together with the discussions above, we expect that the physical phenomenology for all the considered cases here is similar to each other. We therefore only use the classical case in this work to study the impacts on observations and the corresponding constraints on our model. At the perturbation level, we treat the kinetic equilibrium case and free-streaming case differently, which is discussed in Sec.\,\ref{section:perturbation-hb}.

\section{Details of the estimation of the constraint from big bang nuclear synthesis element abundance}\label{section:constraint-BBN-detail}
We use the kinetic equilibrium case to calculate the extra radiation energy density at early times. However, since we have seen that the background evolution in the free-streaming case is similar to that of the kinetic equilibrium case, we expect that the constraint from BBN is also similar for both cases. In the kinetic equilibrium case, the extra radiation energy density in the HiDM scenario is the energy density of the relativistic HiDM. With Eqs.\,\eqref{eq:HiDM-energy-relativistic} and \eqref{eq:x-relativistic-limit} we have,
\begin{equation}\label{eq:HiDM-extra-relativistic-energy}
    \Delta\rho_{\rm{rad}}=\lim_{x\rightarrow0}\rho_{\rm{h}} =\frac{gm^4}{2\pi^2\hbar^3}\frac{F_4^\mp(\Delta_{\rm{ini}})}{C^4}\frac{\hat{a}^4}{a^4}\,,
\end{equation}
where $\Delta_{\rm{ini}}$ is the value of $\Delta$ in the relativistic limit, and the function $F_s^\mp(\Delta)$ is defined in Eq.~\eqref{eq:Fs}. With Eqs.\,\eqref{eq:HiDM-energy-density-non-relativistic} and \eqref{eq:x-nonrelativisitic-limit}, the dark matter energy density in the nonrelativistic phase reads
\begin{equation}\label{eq:HiDM-DM-nonrelativistic-energy}
    \rho_{\rm{dm}}=\lim_{x\rightarrow\infty}\rho_{\rm{h}}=\frac{gm^4}{2\pi\hbar^3}\frac{\sqrt{2}F_{3/2}^\mp(\Delta_{\rm f})}{C^{3/2}}\frac{\hat{a}^3}{a^3}\,.
\end{equation}
The values of $C$, $\Delta_{\rm{ini}}$ and $\Delta_{\rm f}$ for some special cases can be found in Table \ref{tab:transition-parameters}. The above two equations give the extra radiation energy density $\Delta\rho_{\rm{rad}}$ in the early universe,
\begin{equation}\label{eq:HiDM-relativistic-energy-Omegam}
    \Delta\rho_{\rm{rad}}=\frac{F_4^\mp(\Delta_{\rm{ini}})\hat{a}\rho_{\rm{m}}^0}{\sqrt{2}C^{5/2}F_{3/2}^\mp(\Delta_{\rm f})a^4}\simeq \frac{0.8\,\hat{a}\rho_{\rm{m}}^0}{a^4}\,,
\end{equation}
where the approximation is valid for all cases\footnote{This extra radiation energy density is very similar to that when the transition is instantaneous, which would give $\Delta\rho_{\rm{rad}}=\hat{a}\rho_{\rm{m}}^0/a^4$.} (fermionic or bosonic, regardless of $\Delta_{\rm{ini}}$). For cosmological observations, the extra radiation energy density is usually parameterized with $\Delta N_{\rm{eff}}$ as \cite{Planck2018-parameter-constraints} 
\begin{equation}\label{eq:radiation-energy-Neff}
    \Delta \rho_{\rm{rad}} = \Delta N_{\rm{eff}} \times\frac{7}{8}\left(\frac{4}{11}\right)^{4/3}\rho_\gamma\,,
\end{equation}
where $\rho_\gamma$ is the energy density of photons. By equating Eqs.\,\eqref{eq:HiDM-relativistic-energy-Omegam} and \eqref{eq:radiation-energy-Neff} and using $R_{\rm{eq}}^{\rm{T}}=a_{\rm T}^w/a_{\rm{eq}}$ as well as the relation between $\hat{a}$ and $T_{\rm h}^0/m_{\rm h}$ we have
\begin{equation}\label{eq:RTeq-Neff-in-appendix}
    R_{\rm{eq}}^{\rm{T}}=0.23\times\frac{\Omega_{\rm{m}}\Omega_{\gamma}}{\Omega_{\rm{dm}}\Omega_{\rm{rad}}}\Delta N_{\rm{eff}}\,,
\end{equation}
where $\Omega_{\rm{m}}$, $\Omega_{\rm{dm}}$, $\Omega_{\gamma}$ and $\Omega_{\rm{rad}}$ are the density parameters for matter, dark matter, photons, and radiation. Note that $\Omega_{\rm{rad}}$ is calculated by treating neutrinos as relativistic particles. As we shall see, other cosmological constraints on $R_{\rm{eq}}^{\rm{T}}$ are much stronger than the constraint from BBN, so we proceed to only give an estimate of the BBN constraint. We fix all $\Omega$'s to the mean values given in Planck 2018 \cite{Planck2018-parameter-constraints} and adopting a rough upper bound $\Delta N_{\rm{eff}}<0.4$ \cite{Yeh:2022heq}, we derive an upper bound on $R_{\rm{eq}}^{\rm{T}}$  of
\begin{equation}\label{eq:BBN-RTeq-upper-bound-in-appendix}
    R_{\rm{eq}}^{\rm{T}}\lesssim0.065\,,
\end{equation}
where the uncertainties of the $\Omega$'s have been ignored.


\section{The minimum mass - Jeans mass scales  at different redshifts}\label{section:HiDM-minimum-halo-mass}
In section \ref{section:HiDM-halo-mass-function}, we used the Press-Schechter formalism to estimate the halo mass function in the HiDM model. Here we calculate the Jeans mass for dark matter due to its residual velocity dispersion \cite{binney2011galactic}
\begin{equation}\label{eq:HiDM-Jeans-mass}
    M_{\rm{J,h}}(a, R^{\rm{T}}_{\rm{eq}})=\frac{2.92\times\big(c_{\rm{s}}(a)\big)^3}{G^{3/2}\big(\rho_{\rm{m}}(a)\big)^{1/2}}=2.1\times10^{4}\,h^{-1}M_{\odot}\times\Big(\frac{R^{\rm{T}}_{\rm{eq}}}{10^{-3}}\Big)^3a^{-\frac{3}{2}}\,,
\end{equation}
where the subscript ``h'' stands for HiDM. We have used Eq.~\eqref{eq:x-nonrelativisitic-limit} with an adiabatic sound speed $c_{\rm{s}}=\big(\frac{5}{3}\frac{T}{m}\big)^{\frac{1}{2}}=\big(\frac{5}{3}\frac{1}{x}\big)^{\frac{1}{2}}=\big(\frac{5}{3C}\big)^{\frac{1}{2}}\frac{\hat{a}}{a}$, and the classical case constant $C$ from Table \ref{tab:transition-parameters}. This Jeans mass is larger at higher redshifts since the residual temperature was higher. Therefore, such an additional suppression effect is more significant at higher redshifts. For $R_{\rm{eq}}^{\rm{T}}=10^{-3}$, $M_{\rm{J,h}}$ is only $2.1\times10^{4}M_{\odot}$ today but is $3.6\times10^{6}M_{\odot}$ at $z=30$. The halo mass function at each redshift should be further suppressed when the halo mass is near the corresponding Jeans mass. However, $M_{\rm{J,h}}$ is much lower than $M_{1/2}$ caused by the suppression of the initial conditions for nonlinear collapse for the same value of $R^{\rm{T}}_{\rm{eq}}$. Therefore, we may neglect this subdominant suppression effect when we calculate the halo mass function in our $\Lambda$HiDM model.

\bibliography{HDMreference}
\bibliographystyle{jhep}

\end{document}